\documentclass[english]{article}
\usepackage[latin9]{inputenc}
\usepackage{geometry}
\usepackage{mathtools}
\usepackage{amstext}
\usepackage{amssymb}
\usepackage{esint}

\usepackage{color}

\makeatletter
\date{}
\providecommand{\tabularnewline}{\\}

\makeatother

\usepackage[english]{babel}

\begin{document}

\title{\bf Gauge Invariant Description of the $SU(2)$ Higgs model: Ward identities
and Renormalization}

{\author{\textbf{D.~Dudal$^{1,2}$}\thanks{david.dudal@kuleuven.be},
\textbf{D.~M.~van Egmond$^3$}\thanks{duifjemaria@gmail.com},
\textbf{I.~F.~Justo$^4$}\thanks{igorfjusto@gmail.com},
\textbf{G.~Peruzzo$^5$}\thanks{gperuzzofisica@gmail.com},
\textbf{S.~P.~Sorella$^5$}\thanks{silvio.sorella@gmail.com}\\\\\
\textit{{\small $^1$ KU Leuven Campus Kortrijk---Kulak, Department of Physics, Etienne Sabbelaan 53 bus 7657, 8500 Kortrijk, Belgium,}} \\
\textit{{\small $^2$ Ghent University, Department of Physics and Astronomy, Krijgslaan 281-S9, 9000 Gent, Belgium}}\\
\textit{{ \small $^3$ Centre de Physique Th\'eorique, CNRS,}}\\
\textit{{ \small Ecole polytechnique, IP Paris, F-91128, Palaiseau, France}}\\
\textit{{\small $^4$ UFF~- ~Universidade Federal Fluminense,}}\\
\textit{{\small  Instituto de F\'{i}sica, Campus da Praia Vermelha,}}\\
\textit{{ \small Avenida General Milton Tavares de Souza s/n,}}\\
\textit{{ \small 24210-346, Niter\'oi, RJ, Brasil}}\\
\textit{{\small $^5$ UERJ -- Universidade do Estado do Rio de Janeiro,}}\\
\textit{{\small Instituto de F\'{i}sica -- Departamento de F\'{\i}sica Te\'orica -- Rua S\~ao Francisco Xavier 524,}}\\
\textit{{\small 20550-013, Maracan\~a, Rio de Janeiro, Brasil}}\\
}

\maketitle

\vspace*{-1cm}
\begin{abstract}
The renormalization properties of two local gauge invariant composite operators $(O,R^a_{\mu})$ corresponding, respectively, to the gauge invariant description of the Higgs particle and of the massive gauge vector boson, are analyzed to all orders in perturbation theory by means of the algebraic renormalization in the  $SU(2)$ Higgs model, with a single scalar in the fundamental representation, when quantized in the Landau gauge in Euclidean space-time. The present analysis generalizes earlier results presented in the case of the $U(1)$ Higgs model. A powerful global Ward identity, related to an exact custodial symmetry, is derived for the first time, with deep consequences at the quantum level. In particular, the gauge invariant vector operators $R^a_{\mu}$ turn out to be the conserved Noether currents of the above-mentioned custodial symmetry. As such, these composite operators do not renormalize, as expressed by the fact that the renormalization $Z$-factors of the corresponding external sources, needed to define the operators $R^a_{\mu}$ at the quantum level, do not receive any quantum corrections. Another consistency feature of our analysis is that the longitudinal component of the two-point correlation function $\langle R^a_\mu(p) R^b_\nu(-p) \rangle$ exhibits only a tree level non-vanishing contribution which, moreover, is momentum independent, being thus not associated to any physical propagating mode. Finally, we point out that the renowned non-renormalization theorem for the ghost-antighost-vector boson vertex in Landau gauge remains true to all orders, also in presence of the Higgs field.

\end{abstract}

\section{Introduction}
In a previous work \cite{Dudal:2020uwb}, we studied the $SU(2)$ Higgs model
with a complex scalar field in the fundamental representation. In particular, we analyzed a set of
 two-point Green's functions of local
gauge invariant composite operators, one scalar $O(x)$ and a triplet $(R^a_\mu(x), a=1,2,3)$ of  vector operators, namely
\begin{eqnarray}
O(x) & = & \frac{1}{2} \left(  2 vh\left(x\right)+h^{2}\left(x\right)+\rho^{a}\rho^{a}\left(x\right)\right)\label{Oop}, \\[2mm]
R_{\mu}^{a}\left(x\right) & = & -\frac{1}{2}\left[\left(v+h\right)\partial_{\mu}\rho^{a}-\rho^{a}\partial_{\mu}h+\varepsilon^{abc}\rho^{b}\partial_{\mu}\rho^{c}-\frac{g}{2}A_{\mu}^{a}\left(v+h\right)^{2}+g\varepsilon^{abc}A_{\mu}^{b}\rho^{c}\left(v+h\right)+\frac{g}{2}A_{\mu}^{a}\rho^{b}\rho^{b}-gA_{\mu}^{b}\rho^{a}\rho^{b}\right],\nonumber \\ \label{R_op}
\end{eqnarray}
where $A^a_\mu$ is the gauge field, $h$ stands for the Higgs field, $\rho^a$ $( a=1,2,3)$ are the Goldstone bosons and $v$ is the vacuum expectation value  of the scalar complex field. These gauge invariant composite operators were first introduced by 't Hooft \cite{tHooft:1980xss} and later on formalized by Fr\"ohlich-Morchio-Strocchi(FMS) \cite {fms1,fms2} in order to describe the Higgs phenomenon in a gauge invariant fashion, see \cite{Maas:2017wzi,Maas:2017xzh,Maas:2018xxu,Maas:2020kda,Sondenheimer:2019idq} for recent accounts on the subject and \cite{Struyve:2011nz,Berghofer:2021ufy} for a more historical account. In the $U(1)$ case, a gauge invariant reformulation of the Higgs model was also proposed in \cite{Higgs:1966ev}, see also \cite{Struyve:2011nz}, but notice that the non-linear field redefinition invoked there misses a (non-trivial) Jacobian at the quantum level, see \cite{Dudal:2021pvw}, which complicates matters. That such care is needed can also be appreciated from the observation that the (classical) reformulation of \cite{Higgs:1966ev,Struyve:2011nz} eventually leads to equivalence with the original model in the unitary gauge, famous for not being renormalizable.

The relevance of the gauge invariant operators $(O, R^a _\mu)$ can be captured by noticing that, from eqs.\eqref{Oop},\eqref{R_op} it follows  that
\begin{eqnarray}
\langle O(p) O(-p)\rangle & \sim &  \langle h(p)h(-p) \rangle_{\rm tree} + \;\;... \;, \nonumber \\
{\cal P}_{\mu\nu} \langle R^a_\mu(p) R^a_\nu(-p) \rangle & \sim &  {\cal P}_{\mu\nu} \langle A^a_\mu(p) A^b_\nu(-p) \rangle_{\rm tree} + \;\;...  \;, \label{rt}
\end{eqnarray}
where ${\cal P}_{\mu\nu} = \left( \delta_{\mu\nu} -\frac{p_\mu p_\nu}{p^2}\right)$ is the transverse projector and where ... denote the higher order loop corrections
\cite{Dudal:2020uwb}. Eqs.~\eqref{rt} show that the two-point functions of $(O, R^a _\mu)$ are related to those of the elementary fields $(h,A^a_\mu)$.

Concerning the quantum corrections, from the one-loop computations reported in \cite{Dudal:2020uwb} by employing 't Hooft $R_\xi$ gauge, it turns out that  $\langle O(p) O(-p)\rangle$ and ${\cal P}_{\mu\nu} \langle R^a_\mu(p) R^a_\nu(-p) \rangle$ are independent of the gauge parameter $\xi$, while sharing the same pole-mass of the corresponding elementary correlators  $\langle h(p)h(-p) \rangle $ and ${\cal P}_{\mu\nu} \langle A^a_\mu(p) A^b_\nu(-p) \rangle$. It is worth reminding here that the independence of the pole masses from the gauge parameter $\xi$ is in fact a consequence of the so-called Nielsen identities \cite{Nielsen:1975fs, Gambino:1998ec, Gambino:1999ai}. Moreover, both $\langle O(p) O(-p)\rangle$  and ${\cal P}_{\mu\nu} \langle R^a_\mu(p) R^a_\nu(-p) \rangle$ exhibit a suitably subtracted K{\"a}ll{\'e}n-Lehmann representation \cite{weinberg} with positive spectral densities. These features make apparent the fact that the composite operators $(O,R^a_\mu)$ provide a fully consistent gauge invariant setup for the Higgs particle as well as for the vector gauge boson. These results are at odds with those obtained for the elementary correlators, $\langle h(p)h(-p) \rangle$,  $\langle A^a_\mu(p) A^b_\nu(-p) \rangle$, which display an explicit dependence from the gauge parameter $\xi$ as well as their spectral densities which, for some values of $\xi$, are found to violate  positivity, see \cite{Dudal:2020uwb}.

So far, the investigation of the properties of the composite operators $(O,R^a_\mu)$ in the $SU(2)$ Higgs model remains limited at the one-loop order, having not yet reached the all order (or even exact) status achieved in the case of the $U(1)$ Higgs model \cite{Dudal:2019aew,Dudal:2019pyg,Capri:2020ppe,Dudal:2021pvw}, where the analogous of the $SU(2)$ operators, denoted by $(O, V_\mu)$ in \cite{Dudal:2019aew,Dudal:2019pyg,Capri:2020ppe,Dudal:2021pvw}, were shown to obey a set of powerful Ward identities. In particular, in \cite{Capri:2020ppe,Dudal:2021pvw},
we were able to identify the $U(1)$ vector operator $V_\mu$ as the Noether conserved current of the global $U(1)$ symmetry of the model. This feature has led to a powerful Ward identity showing that the operator $V_\mu$ does not get renormalized, a result consistent with $V_\mu$ being a Noether current. Moreover, we also showed that the longitudinal component of $\langle V_\mu(p) V_\nu(-p)\rangle$ does not receive any quantum correction beyond the tree level one, which is completely independent from the momentum $p^2$.  The momentum independence of the longitudinal component of $\langle V_\mu(p) V_\nu(-p)\rangle $ is in fact a necessary condition for a consistent  description of a physical vector massive particle \cite{Itzykson:1980rh}. Let us underline that, in the $U(1)$ case, the above-mentioned results hold to all orders, having been established by means of the algebraic renormalization framework \cite{Becchi:1975nq,Becchi:1974md,Becchi:1974xu,Piguet:1995er}.

The aim of the present paper is to fill the gap between the $U(1)$ and the $SU(2)$ case, investigating the properties of the operators $(O,R^a_\mu)$ to all orders. As we shall see, most of the features established in the $U(1)$ case generalize to $SU(2)$.

More precisely:
\begin{itemize}
\item a set of Ward identities can be established when the composite operators $(O,R^a_\mu)$ are included in the starting action by means of a suitable set of external sources. These Ward identities enable us to prove the  renormalizability of $(O,R^a_\mu)$ to all orders in perturbation theory;

\item similarly to the $U(1)$ case, the gauge invariant vector operators $R^a_\mu$ are the conserved Noether currents of a global custodial exact symmetry displayed by the $SU(2)$ Higgs model. This relevant observation will have deep consequences at the quantum level, implying  the all order non-renormalization of the currents $R^a_\mu$, in agreement with their conserved nature;

\item as happens in the $U(1)$ case \cite{Capri:2020ppe,Dudal:2021pvw}, the longitudinal component of the two-point correlation function  $\left\langle R_{\mu}^{a}(p)R_{\nu}^{b}(-p) \right\rangle $ can be proven to not receive any quantum correction to all orders beyond the tree level contribution which, moreover, is momentum independent. As such, the longitudinal component of $\left\langle R_{\mu}^{a}(p)R_{\nu}^{b}(-p) \right\rangle $ is not associated to any propagating mode;

\item the non-renormalization theorem of the ghost-antighost-gauge boson vertex of  the Landau gauge \cite{Taylor:1971ff,Blasi:1990xz}, which plays a key  role in non-perturbative analyses such as the Schwinger-Dyson  setup \cite{Alkofer:2000wg,Binosi:2009qm,Huber:2018ned}, remains true in presence of the Higgs field.

\end{itemize}
The paper is organized as follows. In Section \ref{HA}, we review briefly the particular $SU(2)$ Higgs model and its BRST quantization in the Landau gauge.   In Section \ref{3} we present a study of the operators $(O,R^a_\mu)$  in terms of the BRST cohomology in order to identify other possible operators with the same quantum numbers which could mix to them at the quantum level. After that, a tree level action including all needed composite operators and related external sources will be written down. Such an action will be taken as the starting point for the quantum analysis of the model to all orders.  In Section \ref{4},  we discuss the exact custodial symmetry, showing that the vector operators $(R_{\mu}^a)$ are nothing but the corresponding conserved Noether currents, a feature which will be translated into a quite powerful Ward identity.   Section \ref{sec:Symmetries-and-Ward}  collects the whole set of Ward identities displayed by the tree level action.  Section \ref{6} is devoted to the all orders algebraic proof of the renormalizability of the model. We shall characterize the most general local invariant counterterm compatible with all Ward identities and we shall show that it can be reabsorbed into the starting action by a redefinition of the fields, parameters and external sources.  In Section \ref{7} we look at the longitudinal part of the two-point correlation function  of the vector operators $(R_{\mu}^a)$ by showing that it does not get any quantum correction beyond a momentum independent tree level one. In Section \ref{8} we present our conclusion and perspectives.

\section{The $SU(2)$ Higgs model and its BRST quantization \label{HA}}

The Euclidean action of the $SU(2)$ Higgs model  with a complex scalar field $\varphi$ in the fundamental representation of the gauge group reads
\begin{eqnarray}
S_{\textrm{Higgs}} & = & \int d^{4}x\left[\frac{1}{4}F_{\mu\nu}^{a}F_{\mu\nu}^{a}+\left(D_{\mu}\varphi\right)^{\dagger}\left(D_{\mu}\varphi\right)+\lambda\left(\varphi^{\dagger}\varphi-\frac{v^{2}}{2}\right)^{2}\right],\label{eq:so}
\end{eqnarray}
with the field strength $F_{\mu\nu}^{a}$  given by
\begin{eqnarray}
F_{\mu\nu}^{a} & = & \partial_{\mu}A_{\nu}^{a}-\partial_{\nu}A_{\mu}^{a}+g\varepsilon^{abc}A_{\mu}^{b}A_{\nu}^{c},\label{eq:F}
\end{eqnarray}
and the covariant derivative
\begin{eqnarray}
D_{\mu}\varphi & = & \partial_{\mu}\varphi-i\frac{g}{2}\tau^{a}A_{\mu}^{a}\varphi,\label{eq:D}
\end{eqnarray}
where $\tau^{a}$ ($a=1,2,3$) are the Pauli matrices and $\varepsilon^{abc}$
the Levi-Civita symbols. The theory has two  coupling constants,
namely, the gauge coupling $g$  and the quartic self-coupling of the scalar field $\lambda$. The  massive parameter $v$ stands for the vacuum  expectation value of $\varphi$.
The action $S_{\textrm{Higgs}}$ is invariant under the gauge tranformations
\begin{eqnarray}
A_{\mu} & \to & UA_{\mu}U{}^{\dagger}+\frac{1}{ig}\left(\partial_{\mu}U\right)U{}^{\dagger}\,,\quad
\varphi ~\to~ U\varphi\,,\quad
\varphi^{\dagger} ~\to~ \varphi{}^{\dagger}U^{\dagger},\label{eq:gauge_gauge-1}
\end{eqnarray}
where $U=\exp\left(-ig\frac{\tau^{a}}{2}\theta^{a}\right)\in SU\left(2\right)$
and $\theta^{a}$ are local parameters. Since we are working in the fundamental
representation of $SU(2)$, we can adopt the following  convenient  parametrization  for the scalar field
\begin{eqnarray}
\varphi\left(x\right) & = & \frac{1}{\sqrt{2}}\left(\begin{array}{c}
\pi\left(x\right)+i\rho^{3}\left(x\right)\\
-\rho^{2}\left(x\right)+i\rho^{1}\left(x\right)
\end{array}\right)
  =  \frac{1}{\sqrt{2}}\left(\pi\left(x\right)I+i\rho^{a}\left(x\right)\tau^{a}\right)\left(\begin{array}{c}
1\\
0
\end{array}\right)\label{eq:Parametrization_phi1}
\end{eqnarray}
where $\pi$, $\rho^{1}$, $\rho^{2}$ and $\rho^{3}$ are real scalar
fields. Looking at the Higgs potential
\begin{eqnarray}
V\left(\varphi\right) & = & \lambda\left(\varphi^{\dagger}\varphi-\frac{v^{2}}{2}\right)^{2}\label{eq:Higgs_potential}
\end{eqnarray}
one can see that its absolute (classical) minimum occurs when $\left|\varphi\right|^{2}=\frac{v^{2}}{2}$.
Choosing  the representative minimum configuration as  $\varphi_{o}=\frac{1}{\sqrt{2}}\left(\begin{array}{c}
v\\
0
\end{array}\right)$, one can consider $\varphi-\varphi_{o}$ as the truly propagating
degree of freedom, which leads to
\begin{eqnarray}
\varphi\left(x\right) & = & \frac{1}{\sqrt{2}}\left(\left(v+h\left(x\right)\right)I+i\rho^{a}\left(x\right)\tau^{a}\right)\left(\begin{array}{c}
1\\
0
\end{array}\right),\label{eq:parametrization_phi2}
\end{eqnarray}
where $h\left(x\right)=\pi\left(x\right)-v$. Rewriting (\ref{eq:so})
in terms of $h\left(x\right)$ and $\rho^{a}\left(x\right)$, one
finds
\begin{eqnarray}
S_{\textrm{Higgs}} & = & \int d^{4}x\left\{ \frac{1}{4}F_{\mu\nu}^{a}F_{\mu\nu}^{a}+\lambda v^{2}h^{2}+\lambda vh^{3}+\lambda vh\rho^{a}\rho^{a}\right.\nonumber \\
 &  & +\frac{1}{4}\lambda h^{4}+\frac{1}{2}\lambda h^{2}\rho^{a}\rho^{a}+\frac{1}{4}\lambda\left(\rho^{a}\rho^{a}\right)^{2}\nonumber \\
 &  & +\frac{1}{2}\left(\partial_{\mu}h\right)^{2}+\frac{1}{2}\left(\partial_{\mu}\rho^{a}\right)^{2}\nonumber \\
 &  & +\frac{1}{2}gA_{\mu}^{a}\rho^{a}\left(\partial_{\mu}h\right)-\frac{1}{2}g\left(v+h\right)A_{\mu}^{a}\left(\partial_{\mu}\rho^{a}\right)+\frac{1}{2}g\varepsilon^{abc}A_{\mu}^{a}\rho^{b}\left(\partial_{\mu}\rho^{c}\right)\nonumber \\
 &  & \left.+\frac{1}{8}g^{2}A_{\mu}^{a}A_{\mu}^{a}\left(v+h\right)^{2}+\frac{1}{8}g^{2}A_{\mu}^{a}A_{\mu}^{a}\rho^{b}\rho^{b}\right\} .\label{eq:so_rewri}
\end{eqnarray}
Looking at (\ref{eq:so_rewri}), we can see all the features of the
Higgs mechanism: the gauge field $A_{\mu}^{a}\left(x\right)$ and the Higgs field $h\left(x\right)$
have acquired masses given by
\begin{eqnarray}
m & = & \frac{1}{2}gv,\quad m_{h}=\sqrt{2\lambda}v\label{eq:masses}
\end{eqnarray}
respectively, while the Goldstone fields $\rho^{a}\left(x\right)$ remain  massless.

\subsection{Gauge Fixing and BRST symmetry}
In order to quantize the theory, we shall adopt
the Landau gauge, i.e.~$\partial_{\mu}A_{\mu}=0$. For the corresponding Faddeev-Popov term we have
\begin{eqnarray}
S_{\textrm{gf}} & = & \int d^{4}x\left[ib^{a}\partial_{\mu}A_{\mu}^{a}+\overline{c}^{a}\partial_{\mu}D_{\mu}^{ab}c^{b}\right],\label{eq:gaugefixing}
\end{eqnarray}
where $b^{a}$ is the Nakanishi-Lautrup field implementing the transversality condition, $\partial_{\mu}A_{\mu}=0$, and $(c^{a},\overline{c}^{a})$ are the
the ghost and  anti-ghost fields. For the gauge fixed action we thus get
\begin{eqnarray}
S & = & S_{\textrm{Higgs}}+S_{\textrm{gf}}\;. \label{eq:action}
\end{eqnarray}
As is well known, expression \eqref{eq:action}  is left invariant by the nilpotent
BRST transformations \cite{Becchi:1975nq,Becchi:1974md,Becchi:1974xu}:
\begin{eqnarray}
sA_{\mu}^{a} & = & -D_{\mu}^{ab}c^{b},\nonumber \\
sc^{a} & = & \frac{g}{2}\varepsilon^{abc}c^{b}c^{c},\nonumber \\
s\overline{c}^{a} & = & ib^{a},\nonumber \\
sb^{a} & = & 0,\nonumber \\
sh & = & \frac{g}{2}c^{a}\rho^{a},\nonumber \\
s\rho^{a} & = & -\frac{g}{2}c^{a}\left(v+h\right)+\frac{g}{2}\varepsilon^{abc}c^{b}\rho^{c},\nonumber \\
sv & = & 0\label{eq:brst_tranformations}
\end{eqnarray}
with
\begin{equation}
s S =0 \;, \quad s^2=0 \;. \label{ss}
\end{equation}

\section{Introduction of the gauge invariant  composite operators $(O(x), R^a_\mu(x))$  \label{3}}
\subsection{The scalar operator $O(x)$}
In order to achieve a better understanding of the gauge invariant local composite operators $(O(x), R^a_\mu(x))$, eqs.\eqref{Oop}, \eqref{R_op}, let us look at them from the viewpoint of the cohomology \cite{Piguet:1995er} of the BRST operator $s$. Let us begin with the scalar operator $O$.

From expression \eqref{Oop}, one observes that the operator $O$ has dimension two. Let us find out the  solution of the cohomology equation
\begin{equation}
s \Delta(x) = 0 \;, \label{Ocoh}
\end{equation}
where $\Delta(x)$ is the most general local colorless scalar polynomial of dimension two in the fields $(A^a_\mu, h, \rho^a, b^a, c^a,\overline{c}^{a}) $ and in the parameter $v$ with vanishing ghost number.  It is not difficult to check out that the most general solution of eq.~\eqref{Ocoh} is given by
\begin{eqnarray}
\Delta\left(x\right) & = & b_{1}O\left(x\right)+b_{2}v^{2}\label{eq:scalar_cohomology}
\end{eqnarray}
with $b_{1}$ and $b_{2}$ arbitrary constants and $O \neq s{\hat O}$ for some local field polynomial ${\hat O}$. Equation \eqref{eq:scalar_cohomology} shows that, apart from the constant term $v^2$, the operator $O$ is the unique term belonging to the cohomology of the BRST operator in the class of the colorless field polynomials of dimension two and with vanishing ghost number.  Let us also notice that, in terms of the complex scalar field $\varphi$, the operator $O$ can be re-written as
\begin{eqnarray}
O & = & \varphi^{\dagger}\varphi-\frac{v^{2}}{2} \;, \label{eq:scalar_op}
\end{eqnarray}
from which its gauge invariance is apparent.

\subsection{The vector operators $R_{\mu}^{a}\left(x\right)$}
In order to introduce the gauge invariant vector operators $R^a_\mu(x)$ let us shortly recall 't Hooft's original construction \cite{tHooft:1980xss}. The first gauge invariant vector quantity can be obtained from
\begin{eqnarray}
O_{\mu}^{3}  =  \varphi^{\dagger}D_{\mu}\varphi \;.
 \label{eq:O_3}
\end{eqnarray}
Following \cite{tHooft:1980xss}, the remaining two operators can be constructed as
\begin{eqnarray}
O_{\mu}^{+} & = & \varphi^{T}\left(\begin{array}{cc}
0 & 1\\
-1 & 0
\end{array}\right)D_{\mu}\varphi,\nonumber \\
O_{\mu}^{-} & = & \left(O_{\mu}^{+}\right)^{\dagger},\label{eq:O_+_-}
\end{eqnarray}
the gauge invariance of which easily follows from the group properties of $SU(2)$. The operators
$\left\{ O_{\mu}^{3},\:O_{\mu}^{+},\:O_{\mu}^{-}\right\} $ yield thus  a set
of three independent gauge invariant vector quantities  with dimension three. The operators $R^a_\mu$, $ a=1,2,3$, can now be obtained  out of $\left\{ O_{\mu}^{3},\:O_{\mu}^{+},\:O_{\mu}^{-}\right\} $ as:
\begin{eqnarray}
R_{\mu}^{1} & = & \frac{i}{2}\left(O_{\mu}^{+}-O_{\mu}^{-}\right),\nonumber \\
R_{\mu}^{2} & = & \frac{1}{2}\left(O_{\mu}^{+}+O_{\mu}^{-}\right),\nonumber \\
R_{\mu}^{3} & = & O_{\mu}^{3}-\frac{i}{2}\partial_{\mu}O \;.\label{eq:R_compon}
\end{eqnarray}
The operators $R^1_\mu$ and $R^2_\mu$ are simple combinations of $(O_{\mu}^{+},O_{\mu}^{-})$, while the operator $R^3_\mu$ is obtained from $O^3_\mu$ by subtracting the divergence of the scalar operator $O(x)$, eq.~\eqref{Oop}. As $O$ is gauge invariant, it turns out that $R^3_\mu$ is  as well. Putting all together, we end up with  the gauge invariant expressions $\{R^a_\mu\}$ given in equation \eqref{R_op}, namely
\begin{eqnarray}
R_{\mu}^{a} & = & -\frac{1}{2}\left[\left(v+h\right)\partial_{\mu}\rho^{a}-\rho^{a}\partial_{\mu}h+\varepsilon^{abc}\rho^{b}\partial_{\mu}\rho^{c}-\frac{g}{2}A_{\mu}^{a}\left(v+h\right)^{2}+g\varepsilon^{abc}A_{\mu}^{b}\rho^{c}\left(v+h\right)+\frac{g}{2}A_{\mu}^{a}\rho^{b}\rho^{b}-gA_{\mu}^{b}\rho^{a}\rho^{b}\right] \;. \nonumber \\ \label{eq:R_op}
\end{eqnarray}
It is worth underlining that the index $a=1,2,3$ in eq.~\eqref{eq:R_op} can be associated to the adjoint representation of $SU(2)$. In fact, as we shall see in the next section, the operators $\{R^a_\mu\}$ transform as a  triplet when both $(A^a_\mu, \rho^a)$ undergo a global transformation in the adjoint representation of $SU(2)$ under which the Higgs field $h$ is a singlet, i.e.~it is left invariant.

As done for the scalar operator $O$, let us have a look  at the vector operators $R^a_\mu$ in terms of the cohomology of the BRST operator, amounting to solve the equation
\begin{equation}
s \Delta_{\mu}^{a}(x) = 0 \;, \label{vcoh}
\end{equation}
where $\Delta_{\mu}^{a}(x)$ is now a local polynomial in the fields and in the parameter $v$ with dimension three and vanishing ghost number. After a rather lengthy algebraic calculation, it turns out that the most general solution of equation \eqref{vcoh} can be written as
\begin{eqnarray}
\Delta_{\mu}^{a}\left(x\right) & = & c_{1}R_{\mu}^{a}+s\left(c_{2}\varepsilon^{abc}A_{\mu}^{b}\overline{c}^{c}-c_{3}i\partial_{\mu}\overline{c}^{a}\right)\nonumber \\
 & = & c_{1}R_{\mu}^{a}+c_{2}\left(-\varepsilon^{abc}\left(D_{\mu}^{bd}c^{d}\right)\overline{c}^{c}+i\varepsilon^{abc}A_{\mu}^{b}b^{c}\right)+c_{3}\left(\partial_{\mu}b^{a}\right),\label{eq:vector_cohomology}
\end{eqnarray}
where $c_{1}$, $c_{2}$ and $c_{2}$ are arbitrary constants and $R^a_\mu \neq s{\hat R}^a_\mu$, for some local polynomial ${\hat R}^a_\mu$. This eq.~\eqref{eq:vector_cohomology} has a deep meaning. It shows that, apart from the BRST exact terms, $s\left(c_{2}\varepsilon^{abc}A_{\mu}^{b}\overline{c}^{c}-c_{3}i\partial_{\mu}\overline{c}^{a}\right)$, the operators $R^a_\mu$ are the unique non-trivial elements of the cohomology of the BRST operator in the sector of the field vector polynomials with dimension three and vanishing ghost number. Since $R^a_\mu$ do depend on neither the Faddeev-Popov ghosts $(c^a,\overline{c}^{a}) $ nor the auxiliary field $b^a$, this is equivalent to state that $R^a_\mu$ are the unique  vector gauge invariant local composite operators of dimension three, a feature which will have several consequences at the quantum level. Of course, the same statement holds true for the scalar operator $O$, eq.~\eqref{Oop}, which, apart from the constant quantity $v^2$, is the unique scalar local gauge invariant field polynomial of dimension two.

In short, identifying the physical observables with the non-trivial elements of the BRST cohomology of ghost number zero, we have identified a physical representation of the scalar and vector degrees of freedom.

\section{The vector operators $R_{\mu}^{a}$ as the Noether currents of the custodial symmetry \label{4}}
This section is devoted to the analysis of the nature of the vector operators $\{ R^a_\mu(x)\}$ which, as we shall see, turn out to be the Noether currents of an exact global symmetry of the action $S$, eq.~\eqref{eq:action}.

More precisely, let us consider the  transformations:
\begin{eqnarray}
\delta^{\mathcal{C}}A_{\mu}^{a} & = & g\varepsilon^{abc}\omega^{b}A_{\mu}^{c},\nonumber \\
\delta^{\mathcal{C}}h & = & 0,\nonumber \\
\delta^{\mathcal{C}}\rho^{a} & = & g\varepsilon^{abc}\omega^{b}\rho^{c},\nonumber \\
\delta^{\mathcal{C}}c^{a} & = & g\varepsilon^{abc}\omega^{b}c^{c},\nonumber \\
\delta^{\mathcal{C}}\overline{c}^{a} & = & g\varepsilon^{abc}\omega^{b}\overline{c}^{c},\nonumber \\
\delta^{\mathcal{C}}b^{a} & = & g\varepsilon^{abc}\omega^{b}b^{c},\label{eq:custodial_transf}
\end{eqnarray}
with $\omega^{a}$ a constant parameter. Eqs.~\eqref{eq:custodial_transf} have a rather transparent meaning: all fields $(A^a_\mu, \rho^a, b^a, c^a, {\overline c}^a) $ undergo an adjoint $SU(2)$ transformation with the Higgs field $h$  being a singlet. It is almost immediate to realize that the transformations \eqref{eq:custodial_transf} yield an exact symmetry of the action $S$
\begin{equation}
\delta^{\mathcal{C}} S = 0 \;. \label{dcst}
\end{equation}
We shall refer to eq.~\eqref{dcst} as the \emph{custodial
symmetry}, see \cite{Dudal:2020uwb}. An important feature of the transformations \eqref{eq:custodial_transf} is expressed by
\begin{eqnarray}
\left[s,\delta^{\mathcal{C}}\right] & = & 0\;, \quad  \left\{ s,\ldots\right\}   \neq  \delta^{\mathcal{C}} \;, \label{eq:comm_s_custodial}
\end{eqnarray}
which tell us that $\delta^{\mathcal{C}}$ commutes with the BRST operator $s$, while it cannot be obtained as the anti-commutator between $s$ and another suitable operator. When translated in terms of Noether currents, eqs.~\eqref{eq:comm_s_custodial} imply that the conserved currents associated to $\delta^{\mathcal{C}}$ belong to the cohomology of the BRST operator $s$, i.e.~the currents are given by BRST invariant local operators which cannot be written in a BRST exact fashion. 

The relevance of the eqs.~\eqref{eq:comm_s_custodial} can be captured by observing that the action $S$ is left invariant by a second set of global transformations:
\begin{eqnarray}
\delta^{\mathcal{R}}A_{\mu}^{a} & = & g\varepsilon^{abc}\omega^{b}A_{\mu}^{c},\nonumber \\
\delta^{\mathcal{R}}h & = & \frac{1}{2}g\omega^{a}\rho^{a},\nonumber \\
\delta^{\mathcal{R}}\rho^{a} & = & \frac{1}{2}g\omega^{b}\left(-\left(v+h\right)\delta^{ab}+\varepsilon^{abc}\rho^{c}\right),\nonumber \\
\delta^{\mathcal{R}}c^{a} & = & g\varepsilon^{abc}\omega^{b}c^{c},\nonumber \\
\delta^{\mathcal{R}}\overline{c}^{a} & = & g\varepsilon^{abc}\omega^{b}\overline{c}^{c},\nonumber \\
\delta^{\mathcal{R}}b^{a} & = & g\varepsilon^{abc}\omega^{b}b^{c},\label{eq:R_transf}
\end{eqnarray}
with
\begin{equation}
\delta^{\mathcal{R}} S = 0 \;, \label{rinv}
\end{equation}
which we shall call \emph{$\mathcal{R}$-symmetry}. Nevertheless, unlike the operator $\delta^{\mathcal{C}}$, eq.~\eqref{eq:custodial_transf}, it can be checked that
\begin{eqnarray}
\left[s,\delta^{\mathcal{R}}\right]  =  0 \;, \quad \left\{ \delta^{\mathcal{G}},s\right\}   =  \delta^{\mathcal{R}} \; ,\label{eq:comm_brst_ghost}
\end{eqnarray}
where $\delta^{\mathcal{G}}$ is given by
\begin{eqnarray}
\delta^{\mathcal{G}}A_{\alpha}^{a} & = & 0,\nonumber \\
\delta^{\mathcal{G}}h & = & 0,\nonumber \\
\delta^{\mathcal{G}}\rho^{a} & = & 0,\nonumber \\
\delta^{\mathcal{G}}c^{a} & = & \omega^{a},\nonumber \\
\delta^{\mathcal{G}}\overline{c}^{a} & = & 0,\nonumber \\
\delta^{\mathcal{G}}b^{a} & = & i\varepsilon^{abc}\omega^{b}\overline{c}^{c},\;\label{eq:ghost_transf}
\end{eqnarray}
and
\begin{equation}
\delta^{\mathcal{G}} S = 0 \;. \label{dgs}
\end{equation}
Therefore, unlike the Noether currents associated to $\delta^{\mathcal{C}}$, those corresponding to the \emph{$\mathcal{R}$-symmetry} will be expressed as an exact BRST variation, as such they are cohomologically trivial and dot no describe observable excitations, in fact leading to zero norm states.

Let us turn now to the explicit  computation  of the Noether currents of the custodial symmetry. To that end let us rewrite  equation \eqref{dcst} as
\begin{eqnarray}
\int d^{4}x\,\mathcal{C}^{a}\left(x\right)S & = & 0 \;, \label{eq:custodial_inv_action}
\end{eqnarray}
where $ \mathcal{C}^{a}\left(x\right)$ stands for the \emph{local} operator
\begin{eqnarray}
\mathcal{C}^{a}\left(x\right) & = &- g\varepsilon^{abc}\left(A_{\mu}^{c}\left(x\right)\frac{\delta}{\delta A_{\mu}^{b}\left(x\right)}+\rho^{c}\left(x\right)\frac{\delta}{\delta\rho^{b}\left(x\right)}+c^{c}\left(x\right)\frac{\delta}{\delta c^{b}\left(x\right)}+\overline{c}^{c}\left(x\right)\frac{\delta}{\delta\overline{c}^{b}}+b^{c}\left(x\right)\frac{\delta}{\delta b^{b}\left(x\right)}\right) \;.\label{eq:custodial_ope}
\end{eqnarray}
According to Noether's theorem, the custodial invariance of $S$ implies
that
\begin{eqnarray}
\mathcal{C}^{a}\left(x\right)S & = & \partial_{\mu}\left(J^{\mathcal{C}}\right)_{\mu}^{a}\left(x\right).\label{eq:noether_custodial}
\end{eqnarray}
The output of a direct calculation reads
\begin{eqnarray}
\left(J^{\mathcal{C}}\right)_{\mu}^{a} & = & gR_{\mu}^{a}-\frac{\delta S}{\delta A_{\mu}^{a}}-s\left(D_{\mu}^{ab}\overline{c}^{b}\right) \;, \label{eq:current_custodial}
\end{eqnarray}
from which we learn that,  modulo equations of
motion and a BRST exact term, the local operators $R_{\mu}^{a}$ are, indeed, precisely the Noether currents of the custodial symmetry. We underline that eq.~\eqref{eq:current_custodial} is in perfect agreement with the analysis of the BRST cohomology done in the previous section, according to which the invariant operators
$\{R_{\mu}^{a}\}$ belong to the cohomology of $s$ and cannot be cast in a BRST exact form. As we shall see in the following, eqs.~\eqref{eq:noether_custodial} and
\eqref{eq:current_custodial} can be translated into a local powerful Ward identity, which will result into strong constraints on the quantum correlation functions, including the renormalizability properties of $R_{\mu}^{a}$.

\section{Ward Identities\label{sec:Symmetries-and-Ward}}
\subsection{Dealing with composite operators at the quantum level: identifying the complete tree-level action $\Sigma$}
We are now ready  to start with the all order analysis of the renormalization of the composite operators $(O,R^a_\mu)$. To that aim we remind that,  in order to construct and renormalize the Green functions of the  operators  $(O, R_{\mu}^{a})$, we have to introduce them in the starting action by means of external (local) sources: $J$
and $\Omega_{\mu}^{a}$, respectively. Moreover, following the algebraic formalism reviewed in \cite{Piguet:1995er}, one needs to introduce external fields for the whole set of quantities entering the cohomology of the BRST operator, eqs.~\eqref{eq:scalar_cohomology},\eqref{eq:vector_cohomology}. In particular, in the case of the vector operators $\{R^a_\mu\}$,  we have to take into account the two BRST exact terms $\left(-\varepsilon^{abc}\left(D_{\mu}^{bd}c^{d}\right)\overline{c}^{c}+i\varepsilon^{abc}A_{\mu}^{b}b^{c}\right) $ and $(\partial_{\mu}b^{a})$. Since
\begin{equation}
-\varepsilon^{abc}\left(D_{\mu}^{bd}c^{d}\right)\overline{c}^{c} )
+i\varepsilon^{abc}A_{\mu}^{b}b^{c} = s(\varepsilon^{abc}A_{\mu}^{b}\overline{c}^{c})  \;, \label{exop}
\end{equation}
it can be introduced by means of a BRST doublet\footnote{It is worth reminding here that a pair $(\alpha, \beta)$ is a BRST doublet  if
\begin{equation}
s \alpha = \beta \;, \quad s\beta = 0 \;. \label{db}
\end{equation}
It can be proven that BRST doublets contribute always to the trivial part of the cohomology of the BRST operator $s$, see \cite{Piguet:1995er}.
}  of external sources $(\Upsilon_{\mu}^{a},   \zeta_{\mu}^{a})$, namely
\begin{eqnarray}
s\Upsilon_{\mu}^{a} & = & \zeta_{\mu}^{a}\,,\quad s\zeta_{\mu}^{a} ~=~ 0,\label{eq:doublet_sources}
\end{eqnarray}
so that
\begin{eqnarray}
s\left(\Upsilon_{\mu}^{a}\varepsilon^{abc}A_{\mu}^{b}\overline{c}^{c}\right) & = & \zeta_{\mu}^{a}\varepsilon^{abc}A_{\mu}^{b}\overline{c}^{c}+\Upsilon_{\mu}^{a}\left(-\varepsilon^{abc}\left(D_{\mu}^{bd}c^{d}\right)\overline{c}^{c}+i\varepsilon^{abc}A_{\mu}^{b}b^{c}\right).\label{eq:exact_term}
\end{eqnarray}
On the other hand, the term $\partial_{\mu}b^{a}$ is linear in the quantum fields, so
it can be introduced in a simple way through the external source $\Theta^a_\mu$.

Therefore, for the whole term accounting for all quantities entering the cohomology of the BRST operator with the same quantum numbers as the composite operators $(O, R^a_\mu)$, we have
\begin{eqnarray}
S_{\Delta} & = & \int d^{4}x\left\{ JO+\eta v^{2} +\Omega_{\mu}^{a}R_{\mu}^{a}\right.\nonumber \\
 &  & \left.+\zeta_{\mu}^{a}\varepsilon^{abc}A_{\mu}^{b}\overline{c}^{c}+\Upsilon_{\mu}^{a}\left(-\varepsilon^{abc}\left(D_{\mu}^{bd}c^{d}\right)\overline{c}^{c}+i\varepsilon^{abc}A_{\mu}^{b}b^{c}\right)+i\Theta_{\mu}^{a}\left(\partial_{\mu}b^{a}\right)\right\} .\label{eq:s_delta}
\end{eqnarray}
where the sources $(J(x), \eta(x), \Omega^a_\mu(x), \Theta^a_\mu(x))$ are BRST invariant,~i.e.
\begin{eqnarray}
s\Omega_{\mu}^{a}=s\Theta_{\mu}^{a}=sJ=s\eta=0\;,   &  & \label{eq:brst_sources}
\end{eqnarray}
a feature which guarantees that
\begin{equation}
s S_{\Delta} = 0 \;. \label{ssdt}
\end{equation}
Nevertheless, in addition to the term $S_{\Delta}$, a second term, $S_s$,  accounting for the non-linearity of the BRST transformations of $(A_{\mu}^{a},c^{a}, h,\rho^a) $, eqs.~\eqref{eq:brst_tranformations}, needs to be added
\begin{eqnarray}
S_{s} & = & \int d^{4}x\left[K_{\mu}^{a}\left(sA_{\mu}^{a}\right)+L^{a}\left(sc^{a}\right)+H\left(sh\right)+P^{a}\left(s\rho^{a}\right)\right] \;, \label{eq:non_linear_tranf_action}
\end{eqnarray}
where the external sources $(K^a_\mu,L^a,H,P^a)$ are BRST invariant
\begin{eqnarray}
sK_{\mu}^{a} & = & sL^{a}=sH=sP^{a}=0,\label{eq:brts_transf_sources}
\end{eqnarray}
so that
\begin{equation}
s S_s = 0 \;. \label{sss}
\end{equation}
Summing up  all  pieces, we can introduce the complete tree level action $\Sigma$ which will be taken as the starting point for the quantum analysis, namely
\begin{eqnarray}
\Sigma & = & S+S_{s}+S_{\Delta}\nonumber \\
 & = & \int d^{4}x\left\{ \frac{1}{4}F_{\mu\nu}^{a}F_{\mu\nu}^{a}+\frac{1}{2}m_{h}^{2}h^{2}+\lambda vh^{3}+\lambda vh\rho^{a}\rho^{a}\right.\nonumber \\
 &  & +\frac{1}{4}\lambda h^{4}+\frac{1}{2}\lambda h^{2}\rho^{a}\rho^{a}+\frac{1}{4}\lambda\left(\rho^{a}\rho^{a}\right)^{2}\nonumber \\
 &  & +\frac{1}{2}\left(\partial_{\mu}h\right)^{2}+\frac{1}{2}\left(\partial_{\mu}\rho^{a}\right)^{2}\nonumber \\
 &  & +\frac{1}{2}gA_{\mu}^{a}\rho^{a}\left(\partial_{\mu}h\right)-\frac{1}{2}g\left(v+h\right)\left(\partial_{\mu}\rho^{a}\right)A_{\mu}^{a}+\frac{1}{2}g\varepsilon^{abc}A_{\mu}^{a}\rho^{b}\partial_{\mu}\rho^{c}\nonumber \\
 &  & +\frac{1}{8}g^{2}A_{\mu}^{a}A_{\mu}^{a}\left(v+h\right)^{2}+\frac{1}{8}g^{2}A_{\mu}^{a}A_{\mu}^{a}\rho^{b}\rho^{b}\nonumber \\
 &  & +ib^{a}\partial_{\mu}A_{\mu}^{a}+\overline{c}^{a}\partial_{\mu}D_{\mu}^{ab}c^{b}\nonumber \\
 &  & +K_{\mu}^{a}\left(sA_{\mu}^{a}\right)+L^{a}\left(sc^{a}\right)+H\left(sh\right)+P^{a}\left(s\rho^{a}\right)\nonumber \\
 &  & +JO+\eta v^{2}\nonumber \\
 &  & \left.+\Omega_{\mu}^{a}R_{\mu}^{a}+\Upsilon_{\mu}^{a}\left(-\varepsilon^{abc}\left(D_{\mu}^{bd}c^{d}\right)\overline{c}^{c}+i\varepsilon^{abc}A_{\mu}^{b}b^{c}\right)+\zeta_{\mu}^{a}\varepsilon^{abc}A_{\mu}^{b}\overline{c}^{c}+i\Theta_{\mu}^{a}\partial_{\mu}b^{a}\right\}. \label{eq:starting_action}
\end{eqnarray}
Evidently
\begin{equation}
s \Sigma = 0 \;. \label{sigmsinv}
\end{equation}

\subsection{Ward identities}
The complete tree level action $\Sigma$ obeys a huge number of Ward identities which we enlist below:
\begin{itemize}
\item the Slavnov-Taylor  identity translating at the functional level the BRST invariance of $\Sigma$
\end{itemize}
\begin{eqnarray}
\mathcal{S}\left(\Sigma\right) & = & 0 \;, \label{eq:slavnov_taylor}
\end{eqnarray}
where
\begin{eqnarray}
\mathcal{S}\left(\Sigma\right) & = & \int d^{4}x\left(\frac{\delta\Sigma}{\delta K_{\mu}^{a}}\frac{\delta\Sigma}{\delta A_{\mu}^{a}}+\frac{\delta\Sigma}{\delta L^{a}}\frac{\delta\Sigma}{\delta c^{a}}+ib^{a}\frac{\delta\Sigma}{\delta\overline{c}^{a}}+\frac{\delta\Sigma}{\delta H}\frac{\delta\Sigma}{\delta h}+\frac{\delta\Sigma}{\delta P^{a}}\frac{\delta\Sigma}{\delta\rho^{a}}+\zeta_{\mu}^{a}\frac{\delta\Sigma}{\delta\Upsilon_{\mu}^{a}}\right) \;, \label{eq:slavnov_op}
\end{eqnarray}
\begin{itemize}
\item the $b^a$ Ward identity expressing in functional form the Landau gauge condition
\end{itemize}
\begin{eqnarray}
\frac{\delta\Sigma}{\delta b^{a}} & = & i\partial_{\mu}A_{\mu}^{a}-i\partial_{\mu}\Theta_{\mu}^{a}-i\varepsilon^{abc}A_{\mu}^{b}\Upsilon_{\mu}^{c} \;, \label{eq:b_equation}
\end{eqnarray}
Notice that the r.h.s.~of eq.~\eqref{eq:b_equation} is a linear breaking. As such, it will be not affected by quantum corrections \cite{Piguet:1995er},
\begin{itemize}
\item the anti-ghost equation
\end{itemize}
\begin{eqnarray}
\frac{\delta\Sigma}{\delta\overline{c}^{a}}+\partial_{\mu}\frac{\delta\Sigma}{\delta K_{\mu}^{a}}+\varepsilon^{abc}\Upsilon_{\mu}^{b}\frac{\delta\Sigma}{\delta K_{\mu}^{c}} & = & \varepsilon^{abc}A_{\mu}^{b}\zeta_{\mu}^{c} \;, \label{eq:antighost_equation}
\end{eqnarray}
\begin{itemize}
\item the local linearly broken ghost Ward identity
\end{itemize}
\begin{eqnarray}
\frac{\delta\Sigma}{\delta c^{a}}-g\varepsilon^{abc}i\overline{c}^{b}\frac{\delta\Sigma}{\delta b^{c}}+g\varepsilon^{abc}\Upsilon_{\mu}^{b}\frac{\delta\Sigma}{\delta\zeta_{\mu}^{c}}+g\partial_{\mu}\left(\frac{\delta\Sigma}{\delta\zeta_{\mu}^{a}}\right) & = & \Delta_{cl}^{a} \;, \label{eq:ghost_equation}
\end{eqnarray}
where $\Delta_{cl}^{a} $ stands for the local linear breaking
\begin{eqnarray}
\Delta_{cl}^{a} &=& -\partial^{2}\overline{c}^{a}-D_{\mu}^{ab}K_{\mu}^{b}+g\varepsilon^{bac}L^{b}c^{c}  -g\varepsilon^{abc}\left(\partial_{\mu}\Theta_{\mu}^{c}\right)\overline{c}^{b}\nonumber \\
 &  & -\frac{g}{2}H\rho^{a}+\frac{g}{2}P^{a}\left(v+h\right)-\frac{g}{2}\varepsilon^{bac}P^{b}\rho^{c}
  +\varepsilon^{bac}\partial_{\mu}\left(\Upsilon_{\mu}^{b}\overline{c}^{c}\right)\;. \label{eq:ghost_equation_br}
\end{eqnarray}
It is worth noticing that, unlike the usual ghost Ward identity 
of the Landau gauge \cite{Piguet:1995er,Blasi:1990xz} which is an integrated equation,
 the ghost identity \eqref{eq:ghost_equation} is local, that is non-integrated. In the 
present case, this feature is due to the presence of the external 
sources $(\Upsilon, \zeta)$,

\begin{itemize}
\item the exact $\mathcal{R}$ symmetry
\end{itemize}
\begin{eqnarray}
\mathcal{R}^{a}\left(\Sigma\right) & = & 0 \;, \label{eq:R_equation}
\end{eqnarray}
where
\begin{eqnarray}
\mathcal{R}^{a} & = & g\varepsilon^{abc}A_{\mu}^{b}\frac{\delta}{\delta A_{\mu}^{c}}+g\varepsilon^{abc}K_{\mu}^{b}\frac{\delta}{\delta K_{\mu}^{c}}\nonumber \\
 &  & +g\varepsilon^{abc}c^{b}\frac{\delta}{\delta c^{c}}+g\varepsilon^{abc}L^{b}\frac{\delta}{\delta L^{c}}\nonumber \\
 &  & +g\varepsilon^{abc}\overline{c}^{b}\frac{\delta}{\delta\overline{c}^{c}}+g\varepsilon^{abc}b^{b}\frac{\delta}{\delta b^{c}}+\frac{1}{2}g\rho^{a}\frac{\delta}{\delta h}\nonumber \\
 &  & +\frac{1}{2}gP^{a}\frac{\delta}{\delta H}+\frac{1}{2}g\left(-\delta^{ca}\left(v+h\right)+\varepsilon^{cab}\rho^{b}\right)\frac{\delta}{\delta\rho^{c}}\nonumber \\
 &  & -\frac{1}{2}g\left(H\delta^{ca}-\varepsilon^{abc}P^{b}\right)\frac{\delta}{\delta P^{c}}\nonumber \\
 &  & +g\varepsilon^{abc}\Upsilon_{\mu}^{b}\frac{\delta}{\delta\Upsilon_{\mu}^{c}}+g\varepsilon^{abc}\zeta_{\mu}^{b}\frac{\delta}{\delta\zeta_{\mu}^{c}} \;, \label{eq:R_ope}
\end{eqnarray}
\begin{itemize}
\item the  $c$-$\overline{c}$ Ward identity
\end{itemize}
\begin{eqnarray}
\mathcal{\tau}\left(\Sigma\right) & = & -\frac{1}{g}\int d^{4}x\zeta_{\mu}^{a}\left(\partial_{\mu}c^{a}\right) \;, \label{eq:SL_2C_equation}
\end{eqnarray}
where
\begin{eqnarray}
\mathcal{\tau}\left(\Sigma\right) & = & \int d^{4}x\left\{ c^{a}\frac{\delta\Sigma}{\delta\overline{c}^{a}}-i\frac{\delta\Sigma}{\delta b^{a}}\frac{\delta\Sigma}{\delta L^{a}}+\frac{1}{g}\zeta^{a}_{\mu}\frac{\delta\Sigma}{\delta K^{a}_{\mu}}-\left(\frac{1}{g}\left(\partial_{\mu}\Upsilon_{\mu}^{a}\right)+i\left(\partial_{\mu}\Theta_{\mu}^{a}\right)\right)\frac{\delta\Sigma}{\delta L^{a}}\right\} \;,  \label{eq:SL_2C_op}
\end{eqnarray}
\begin{itemize}
\item the linearly broken integrated equation of the Higgs field $h$
\end{itemize}
\begin{eqnarray}
\int d^{4}x\left(\frac{\delta\Sigma}{\delta h}-2\lambda v\frac{\delta\Sigma}{\delta J}\right)-\frac{\partial\Sigma}{\partial v} & = & \int d^{4}xv\left(J-2\eta\right) \;, \label{eq:h_equation}
\end{eqnarray}
\begin{itemize}
\item the local custodial Ward identity
\end{itemize}
\begin{eqnarray}
\mathcal{C}^{a}\left(\Sigma\right) & = & \frac{1}{4}gv^{2}\partial_{\mu}\Omega_{\mu}^{a}+i\varepsilon^{bac}\partial_{\mu}\left(\Upsilon_{\mu}^{b}b^{c}\right)+\varepsilon^{bac}\partial_{\mu}\left(\zeta_{\mu}^{b}\overline{c}^{c}\right)-ig\varepsilon^{bac}\partial_{\mu}\left(b^{c}\Theta_{\mu}^{b}\right)-i\partial^{2}b^{a}.\label{eq:custodial_eq}
\end{eqnarray}
where

\begin{eqnarray}
\mathcal{C}^{a}\left(\Sigma\right) & = & g\varepsilon^{abc}\left( A_{\mu}^{b}\frac{\delta\Sigma}{\delta A_{\mu}^{c}}+\rho^{b}\frac{\delta\Sigma}{\delta\rho^{c}}+c^{b}\frac{\delta\Sigma}{\delta c^{c}}+\overline{c}^{b}\frac{\delta\Sigma}{\delta\overline{c}^{c}}+b^{b}\frac{\delta\Sigma}{\delta b^{c}}\right. \nonumber \\
 &  & +K_{\mu}^{b}\frac{\delta\Sigma}{\delta K_{\mu}^{c}}+L^{b}\frac{\delta\Sigma}{\delta L^{c}}+P^{b}\frac{\delta\Sigma}{\delta P^{c}}\nonumber \\
 &  & \left. +\Omega_{\mu}^{b}\frac{\delta\Sigma}{\delta\Omega_{\mu}^{m}}+\Upsilon_{\mu}^{b}\frac{\delta\Sigma}{\delta\Upsilon_{\mu}^{c}}+\zeta_{\mu}^{b}\frac{\delta\Sigma}{\delta\zeta_{\mu}^{c}}+\Theta_{\mu}^{b}\frac{\delta\Sigma}{\delta\Theta_{\mu}^{c}} \right) \nonumber \\
 &  & +\partial_{\mu}\frac{\delta\Sigma}{\delta A_{\mu}^{a}}-g\partial_{\mu}\frac{\delta\Sigma}{\delta\Omega_{\mu}^{a}}-\frac{1}{2}g\partial_{\mu}\left(\Omega_{\mu}^{a}\frac{\delta\Sigma}{\delta J}\right)+g\partial_{\mu}\frac{\delta\Sigma}{\delta\Upsilon_{\mu}^{a}} \;, \label{eq:custodial_op}
\end{eqnarray}
\begin{itemize}
\item the external sources $\eta$ and $\Theta$ Ward identities
\end{itemize}
\begin{eqnarray}
\frac{\delta\Sigma}{\delta\eta} & = & v^{2} \;, \label{eq:eta_equation}\\
\frac{\delta\Sigma}{\delta\Theta_{\mu}^{a}} & = & i\partial_{\mu}b^{a} \;. \label{eq:Teta_equation}
\end{eqnarray}

\section{All orders algebraic analysis of the renormalizability \label{6}}

\subsection{Characterization of the local invariant counterterm}
From the power counting, the invariant local counterterm $\Sigma_{\textrm{ct}}$
which can be freely added at any given loop order in perturbation theory, is a local, integrated polynomial of dimension four in the fields, sources and parameters with vanishing ghost number. According to algebraic setup \cite{Piguet:1995er}, the characterization of $\Sigma_{\textrm{ct}}$ is done by requiring that the so-called bare action $\Sigma_{\textrm{bare}}$ defined as
\begin{eqnarray}
 \Sigma+\epsilon\Sigma_{\textrm{ct}} = \Sigma_{\textrm{bare}}+O\left(\epsilon^{2} \right) ,\label{eq:bare_renor}
\end{eqnarray}
where $\epsilon$ is an expansion parameter, satisfies, up to the order $\epsilon^2$, the same Ward identities obeyed by the tree level action $\Sigma$. At the end, one has  to check  that $\Sigma_{\textrm{bare}}$ can be obtained from the tree level action $\Sigma$ by a suitable redefinition of the fields, coupling constants, mass parameters and external sources. Notice that this can amount to allowing mixing between quantities with identical quantum numbers. Requiring thus that $\Sigma_{\textrm{bare}}$ fulfills the same Ward identities of $\Sigma$, it follows that:
\begin{eqnarray}
\mathcal{S}\left(\Sigma_{\textrm{bare}}\right) & = & 0 \;,
\end{eqnarray}
\begin{eqnarray}
\frac{\delta\Sigma_{\textrm{bare}}}{\delta b^{a}} & = & i\partial_{\mu}A_{\mu}^{a}-\partial_{\mu}\Theta_{\mu}^{a}-i\varepsilon^{abc}A_{\mu}^{b}\Upsilon_{\mu}^{c} \;,
\end{eqnarray}
\begin{eqnarray}
\frac{\delta\Sigma_{\textrm{bare}}}{\delta\overline{c}^{a}}+\partial_{\mu}\frac{\delta\Sigma_{\textrm{bare}}}{\delta K_{\mu}^{a}}+\varepsilon^{abc}\Upsilon_{\mu}^{b}\frac{\delta\Sigma_{\textrm{bare}}}{\delta K_{\mu}^{c}} & = & \varepsilon^{abc}A_{\mu}^{b}\zeta_{\mu}^{c} \;,
\end{eqnarray}
\begin{eqnarray}
\int d^{4}x\left[\frac{\delta\Sigma_{\textrm{bare}}}{\delta c^{a}}+g\varepsilon^{abc}\left(-i\overline{c}^{b}\frac{\delta\Sigma_{\textrm{bare}}}{\delta b^{c}}+\Upsilon_{\mu}^{b}\frac{\delta\Sigma_{\textrm{bare}}}{\delta\zeta_{\mu}^{c}}\right)\right] & = & \int d^4x\; \Delta_{cl}^{a} \;,
\end{eqnarray}
\begin{eqnarray}
\tau\left(\Sigma_{\textrm{bare}}\right) & = & -\frac{1}{g}\int d^{4}x\zeta_{\mu}^{a}\left(\partial_{\mu}c^{a}\right) \;,
\end{eqnarray}
\begin{eqnarray}
\int d^{4}x\left(\frac{\delta\Sigma_{\textrm{bare}}}{\delta h}-2\lambda v\frac{\delta\Sigma_{\textrm{bare}}}{\delta J}\right)-\frac{\partial\Sigma_{\textrm{bare}}}{\partial v} & = & \int d^{4}xv\left(J-2\eta\right) \;,
\end{eqnarray}
\begin{eqnarray}
\mathcal{C}^{a}\left(\Sigma_{\textrm{bare}}\right) & = & \frac{1}{4}v^{2}\partial_{\mu}\Omega_{\mu}^{a}+\frac{1}{g}i\varepsilon^{bac}\partial_{\mu}\left(\Upsilon_{\mu}^{b}b^{c}\right)+\frac{1}{g}\varepsilon^{bac}\partial_{\mu}\left(\zeta_{\mu}^{b}\overline{c}^{c}\right)-i\varepsilon^{bac}\partial_{\mu}\left(b^{c}\Theta_{\mu}^{b}\right)-\frac{i}{g}\partial^{2}b^{a} \;,
\end{eqnarray}
\begin{eqnarray}
\frac{\delta\Sigma_{\textrm{bare}}}{\delta\eta} & = & v^{2} \;,
\end{eqnarray}
\begin{eqnarray}
\frac{\delta\Sigma_{\textrm{bare}}}{\delta\Theta_{\mu}^{a}} & = & i\partial_{\mu}b^{a} \;,
\end{eqnarray}
From eq.~(\ref{eq:bare_renor}), the invariant counterterm $\Sigma_{\rm ct}$ is found to obey the following constraints:
\begin{eqnarray}
\frac{\delta\Sigma_{\textrm{ct}}}{\delta b^{a}} & = & 0 \;, \label{eq:b_ct_eq}
\end{eqnarray}
\begin{eqnarray}
\frac{\delta\Sigma_{\textrm{ct}}}{\delta\eta} & = & 0 \;, \label{eq:eta_ct_eq}
\end{eqnarray}
\begin{eqnarray}
\frac{\delta\Sigma_{\textrm{ct}}}{\delta\Theta_{\mu}^{a}} & = & 0\;, \label{eq:teta_ct_eq}
\end{eqnarray}
\begin{eqnarray}
\frac{\delta\Sigma_{\textrm{ct}}}{\delta\overline{c}^{a}}+\partial_{\mu}\frac{\delta\Sigma_{\textrm{ct}}}{\delta K_{\mu}^{a}}+\varepsilon^{abc}\Upsilon_{\mu}^{b}\frac{\delta\Sigma_{\textrm{ct}}}{\delta K_{\mu}^{c}} & = & 0 \;,
\end{eqnarray}
\begin{eqnarray}
\int d^{4}x\left[\frac{\delta\Sigma_{\textrm{ct}}}{\delta c^{a}}+g\varepsilon^{abc}\left(-i\overline{c}^{b}\frac{\delta\Sigma_{\textrm{ct}}}{\delta b^{c}}+\Upsilon_{\mu}^{b}\frac{\delta\Sigma_{\textrm{ct}}}{\delta\zeta_{\mu}^{c}}\right)\right] & = & 0 \;,
\end{eqnarray}
\begin{eqnarray}
\int d^{4}x\left(\frac{\delta\Sigma_{\textrm{ct}}}{\delta h}-2\lambda v\frac{\delta\Sigma_{\textrm{ct}}}{\delta J}\right)-\frac{\partial\Sigma_{\textrm{ct}}}{\partial v} & = & 0 \;,
\end{eqnarray}
\begin{eqnarray}
\mathcal{C}^{a}\left(\Sigma_{\textrm{ct}}\right) & = & 0 \;.
\end{eqnarray}
Due to the non-linearity of the Slavnov-Taylor identity,   it follows that
\begin{eqnarray}
\mathcal{S}\left(\Sigma_{\textrm{bare}}\right) & = & \mathcal{S}\left(\Sigma\right)+\epsilon\mathcal{B}_{\Sigma}\left(\Sigma_{\textrm{ct}}\right)+O\left(\epsilon^{2}\right) \;,
\end{eqnarray}
where
\begin{eqnarray}
\mathcal{B}_{\Sigma} & = & \int d^{4}x\left\{ \frac{\delta\Sigma}{\delta K_{\mu}^{a}}\frac{\delta}{\delta A_{\mu}^{a}}+\frac{\delta\Sigma}{\delta A_{\mu}^{a}}\frac{\delta}{\delta K_{\mu}^{a}}+\frac{\delta\Sigma}{\delta L^{a}}\frac{\delta}{\delta c^{a}}+\frac{\delta\Sigma}{\delta c^{a}}\frac{\delta}{\delta L^{a}}\right.\nonumber \\
 &  & \left.+ib^{a}\frac{\delta}{\delta\overline{c}^{a}}+\frac{\delta\Sigma}{\delta H}\frac{\delta}{\delta h}+\frac{\delta\Sigma}{\delta h}\frac{\delta}{\delta H}+\frac{\delta\Sigma}{\delta P^{a}}\frac{\delta}{\delta\rho^{a}}+\frac{\delta\Sigma}{\delta\rho^{a}}\frac{\delta}{\delta P^{a}}+\zeta_{\mu}^{a}\frac{\delta}{\delta\Upsilon_{\mu}^{a}}\right\}
\end{eqnarray}
is the so-called linearized nilpotent Slavnov-Taylor operator \cite{Piguet:1995er}:
\begin{equation}
\mathcal{B}_{\Sigma} \mathcal{B}_{\Sigma} = 0 \;. \label{nillin}
\end{equation}
Since $\mathcal{S}\left(\Sigma\right)=0$, we have the condition

\begin{eqnarray}
\mathcal{B}_{\Sigma}\left(\Sigma_{\textrm{ct}}\right) & = & 0 \;, \label{eq:slavnov_linear_equation}
\end{eqnarray}
implying that $\Sigma_{\textrm{ct}}$ belongs to the cohomology of the linearized operator $\mathcal{B}_{\Sigma}$ in the space of the integrated local polynomials in the fields and sources with dimension four and vanishing ghost number, see Tables~\ref{tab:Dimensions-and-ghost_fields} and \ref{tab:Dimensions-and-ghost-sources}.
\begin{center}
\begin{table}
\begin{centering}
\begin{tabular}{|c|c|c|c|c|c|c|}
\hline
 & $A_{\mu}^{a}$ & $h$ & $\rho^{a}$ & $c^{a}$ & $\overline{c}^{a}$ & $b^{a}$\tabularnewline
\hline
\hline
dimension & 1 & 1 & 1 & 0 & 2 & 2\tabularnewline
\hline
ghost number & 0 & 0 & 0 & 1 & -1 & 0\tabularnewline
\hline
\end{tabular}
\par\end{centering}
\caption{Mass dimensions and ghost numbers of the fields.\label{tab:Dimensions-and-ghost_fields}}
\end{table}
\par\end{center}
\begin{center}
\begin{table}
\begin{centering}
\begin{tabular}{|c|c|c|c|c|c|c|c|c|c|}
\hline
 & $K_{\mu}^{a}$ & $H$ & $P^{a}$ & $L^{a}$ & $J$ & $\eta$ & $\Upsilon_{\mu}^{a}$ & $\zeta_{\mu}^{a}$ & $\Theta_{\mu}^{a}$\tabularnewline
\hline
\hline
dimension & 3 & 3 & 3 & 4 & 2 & 2 & 1 & 1 & 1\tabularnewline
\hline
ghost number & -1 & -1 & -1 & -2 & 0 & 0 & 0 & 1 & 1\tabularnewline
\hline
\end{tabular}
\par\end{centering}
\caption{Mass dimensions and ghost numbers of the sources.\label{tab:Dimensions-and-ghost-sources}}
\end{table}
\par\end{center}
In order to find out the most general expression for $\Sigma_{\textrm{ct}}$ we start with the condition \eqref{eq:slavnov_linear_equation} which enables us to set
\begin{eqnarray}
\Sigma_{\textrm{ct}} & = & \Delta+\mathcal{B}_{\Sigma}\Delta^{\left(-1\right)}
\;, \label{werg}
\end{eqnarray}
where $\Delta$ and $\mathcal{B}_{\Sigma}\Delta^{\left(-1\right)}$ identify
the non-trivial and trivial cohomology, respectively, of the linearized
Slavnov-Taylor operator $\mathcal{B}_{\Sigma}$. Taking already into account the constraints (\ref{eq:b_ct_eq})-(\ref{eq:teta_ct_eq}) and making use of the general results on the cohomology of non-Abelian gauge theories, see \cite{Piguet:1995er}, we get
\begin{eqnarray}
\Delta & = & \int d^{4}x\left\{ a_{0}\frac{1}{4}F_{\mu\nu}^{a}F_{\mu\nu}^{a}+a_{1}O^{2}+a_{2}v^{2}O\right.\nonumber \\
 &  & +a_{3}JO+a_{4}Jv^{2}+a_{5}v^{4}\nonumber \\
 &  & +a_{6}\Omega_{\mu}^{a}R_{\mu}^{a}+a_{7}\Omega_{\mu}^{a}\Omega_{\mu}^{a}v^{2}+a_{8}\Omega_{\mu}^{a}\Omega_{\mu}^{a}O\nonumber \\
 &  & +a_{9}J^{2}+a_{10}J\Omega_{\mu}^{a}\Omega_{\mu}^{a}+a_{11}\left(\partial_{\mu}\Omega_{\mu}^{a}\right)\left(\partial_{\nu}\Omega_{\nu}^{a}\right)\nonumber \\
 &  & +a_{12}\Omega_{\mu}^{a}\partial^{2}\Omega_{\mu}^{a}+a_{13}\varepsilon^{abc}\Omega_{\mu}^{a}\Omega_{\nu}^{b}\partial_{\mu}\Omega_{\nu}^{c}\nonumber \\
 &  & \left.+a_{14}\Omega_{\mu}^{a}\Omega_{\mu}^{a}\Omega_{\nu}^{b}\Omega_{\nu}^{b}+a_{15}\Omega_{\mu}^{a}\Omega_{\nu}^{a}\Omega_{\mu}^{b}\Omega_{\nu}^{b}\right\}
\end{eqnarray}
and
\begin{eqnarray}
\Delta^{\left(-1\right)} & = & \int d^{4}x\left\{ \overline{c}^{a}\left[d_{1}\partial_{\mu}A_{\mu}^{a}+d_{2}\partial_{\mu}\Omega_{\mu}^{a}+d_{3}\partial_{\mu}\Upsilon_{\mu}^{a}\right.\right.\nonumber \\
 &  & +d_{4}\varepsilon^{abc}A_{\mu}^{b}\Omega_{\mu}^{c}+d_{5}\varepsilon^{abc}A_{\mu}^{b}\Upsilon_{\mu}^{c}+d_{6}\varepsilon^{abc}\Omega_{\mu}^{b}\Upsilon_{\mu}^{c}\nonumber \\
 &  & \left.+d_{7}\varepsilon^{abc}\overline{c}^{b}c^{c}+d_{8}h\rho^{a}+d_{9}v\rho^{a}\right]\nonumber \\
 &  & +K_{\mu}^{a}\left[d_{10}A_{\mu}^{a}+d_{11}\Omega_{\mu}^{a}+d_{12}\Upsilon_{\mu}^{a}\right]\nonumber \\
 &  & +H\left[d_{13}v+d_{14}h\right]\nonumber \\
 &  & \left.+d_{15}P^{a}\rho^{a}+d_{16}L^{a}c^{a}\right\} .
\end{eqnarray}
where $(a_0, ...,a_{15})$ and $(d_1,....,d_{16})$ are free dimensionless parameters.

After imposing all the remaining constraints, a tedious but purely algebraic analysis gives the following results:
\begin{eqnarray}
\Delta & = & \int d^{4}x\left\{ a_{0}\frac{1}{4}F_{\mu\nu}^{a}F_{\mu\nu}^{a}+a_{1}\left(O^{2}+\frac{1}{\lambda}JO+\frac{1}{4\lambda^{2}}J^{2}-\frac{1}{4\lambda}\Omega_{\mu}^{a}\Omega_{\mu}^{a}O-\frac{1}{8\lambda^{2}}J\Omega_{\mu}^{a}\Omega_{\mu}^{a}+\frac{1}{64\lambda^{2}}\Omega_{\mu}^{a}\Omega_{\mu}^{a}\Omega_{\nu}^{b}\Omega_{\nu}^{b}\right)\right.\nonumber \\
 &  & +a_{2}\left(v^{2}O-\frac{1}{\lambda}JO+\frac{1}{2\lambda}Jv^{2}-\frac{1}{2\lambda^{2}}J^{2}-\frac{1}{8\lambda}\Omega_{\mu}^{a}\Omega_{\mu}^{a}v^{2}+\frac{1}{4\lambda^{2}}J\Omega_{\mu}^{a}\Omega_{\mu}^{a}+\frac{1}{4\lambda}\Omega_{\mu}^{a}\Omega_{\mu}^{a}O-\frac{1}{32\lambda^{2}}\Omega_{\mu}^{a}\Omega_{\mu}^{a}\Omega_{\nu}^{b}\Omega_{\nu}^{b}\right)\nonumber \\
 &  & +a_{5}\left(v^{4}-2\frac{1}{\lambda}Jv^{2}+\frac{1}{\lambda^{2}}J^{2}+\frac{1}{2\lambda}\Omega_{\mu}^{a}\Omega_{\mu}^{a}v^{2}-\frac{1}{2\lambda^{2}}J\Omega_{\mu}^{a}\Omega_{\mu}^{a}+\frac{1}{16\lambda^{2}}\Omega_{\mu}^{a}\Omega_{\mu}^{a}\Omega_{\nu}^{b}\Omega_{\nu}^{b}\right)\nonumber \\
 &  & \left.-\frac{1}{2}a_{12}\mathcal{F}_{\mu\nu}^{a}\left(\Omega\right)\mathcal{F}_{\mu\nu}^{a}\left(\Omega\right)\right\} \;, \label{delta_final_nontrivial}
\end{eqnarray}
and
\begin{eqnarray}
\Delta^{\left(-1\right)} & = & \int d^{4}x\left\{ -d_{1}\left(K_{\mu}^{a}+\partial_{\mu}\overline{c}^{a}+\varepsilon^{abc}\Upsilon_{\mu}^{b}\overline{c}^{c}\right)\left(A_{\mu}^{a}-\frac{1}{g}\Upsilon_{\mu}^{a}\right)\right.\nonumber \\
 &  & \left.+d_{13}\left[H\left(v+h\right)+P^{a}\rho^{a}\right]\right\} \;, \label{eq:delta_final}
\end{eqnarray}
where
\begin{eqnarray}
\mathcal{F}_{\mu\nu}^{a}\left(\Omega\right) & = & \partial_{\mu}\Omega_{\nu}^{a}-\partial_{\nu}\Omega_{\mu}^{a}-\varepsilon^{abc}\Omega_{\mu}^{b}\Omega_{\nu}^{c}\,.
\label{worgi1}
\end{eqnarray}
We see thus that the most general final form of the local invariant counterterm $\Sigma_{\textrm{ct}}$ contains seven free parameters, namely: $(a_0, a_1, a_2, a_5, a_{12})$ and $(d_1,d_2)$. In particular, from expression \eqref{delta_final_nontrivial}, one notices the presence of non-linear terms in the BRST invariant external sources $(J,\Omega^a_\mu)$ which are not present in the tree level action $\Sigma$. Nevertheless, these terms, which start from one-loop onward  \cite{Itzykson:1980rh}, are needed to renormalize the two-point correlation functions $\langle O(x) O(y) \rangle$, $\langle R^a_\mu(x) R^b_\nu(y) \rangle$,  defined as
\begin{eqnarray}
\langle O(x) O(y) \rangle & = & \frac{\delta^2 {{Z}_c}}{\delta J(x) \delta J(y)}\Big|_{\rm sources=0} \;, \nonumber \\
\langle R^a_\mu(x) R^b_\nu(y) \rangle & = & \frac{\delta^2 {{ Z}_c}}{\delta \Omega^a_\mu(x) \delta \Omega^b_\nu(y)}\Big|_{\rm sources=0} \;. \label{2-pc}
\end{eqnarray}
where $Z_c$ is the functional generator of the connected Green functions of the model:
\begin{eqnarray}
Z_{c} & = & \Gamma+\sum_{\textrm{fields}\:\phi} \int d^4x J_{\phi}\phi.\label{eq:connected_funct}
\end{eqnarray}
with $\Gamma$ the generator of the $1PI$ Green functions.

Let us also remind that the presence of the BRST invariant counterterm $a_{2}v^{2}O$, which also starts from one-loop onward, is well known in the renormalization of the Higgs model, see \cite{Becchi:1975nq,Kraus:1998ud}. The free coefficient $a_2$ is fixed, order by order in the loop expansion, so as to kill the tadpoles, i.e.~to ensure that $\langle h \rangle =0$. Notice also that the expression \eqref{delta_final_nontrivial} contains the vacuum counterterm $a_{5} v^{4}$ which is allowed by power counting and dimensionality. As discussed in \cite{Dudal:2021pvw}, the parameter $a_5$ can be chosen in such a way that the perturbative dimension two condensate $\langle O \rangle_{\rm pert}$ vanishes order by order: $\langle O \rangle_{\rm pert}=0$. In turn, the Ward identity \eqref{eq:h_equation}, taken together with the two conditions $\langle h \rangle =0$ and $\langle O \rangle_{\rm pert}=0$, then ensures that the (quantum) vacuum energy ${\cal E}_v$ attains its minimum still at $v$ \cite{Dudal:2021pvw}, namely  $\frac{\partial{\cal E}_v}{\partial v}=0$. We shall come back to this point in Section \ref{7}.

Finally, let us spend a few words on the non-linear counterterm $a_{12}\mathcal{F}_{\mu\nu}^{a}\left(\Omega\right)\mathcal{F}_{\mu\nu}^{a}\left(\Omega\right)$ in the external source $\Omega^a_\mu$ showing up in eq.~\eqref{delta_final_nontrivial}. As one can easily figure out, its peculiar form is dictated by the local custodial Ward identity, eqs. \eqref{eq:custodial_eq},\eqref{eq:custodial_op}, according to which both $A^a_\mu$ and $\Omega^a_\mu$ undergo a similar transformation:
\begin{eqnarray}
\delta A^a_{\mu}&=& -\partial_{\mu} \omega^a+g\varepsilon^{abc}\omega ^b A^c_{\mu}, \nonumber \\
\delta \Omega^a_{\mu}&=& -\partial_{\mu} \omega^a+\varepsilon^{abc}\omega ^b \Omega ^c_{\mu},
\end{eqnarray}
which gives rise to the invariant term $\mathcal{F}_{\mu\nu}^{a}\left(\Omega\right)\mathcal{F}_{\mu\nu}^{a}\left(\Omega\right)$.

\subsection{Identifying the bare action and the renormalization $Z$-factors  }
Having characterized the most general local invariant counterterm $\Sigma_{\rm ct}$, eqs.~\eqref{delta_final_nontrivial}, \eqref{eq:delta_final}, compatible with all Ward identities, we can now proceed to write down the bare action $\Sigma_{\textrm{bare}}$ from which the $Z$-factors of all fields, coupling constants, parameters and external sources can be read off. Taking into account the non-linear terms in the external sources needed from one-loop onward as well as the BRST invariant counterterms to enforce the vanishing of the tadpoles, $\langle h \rangle =0$, and of the dimension two perturbative condensate, $\langle O \rangle_{\rm pert}=0$, from eq.~\eqref{eq:bare_renor} for the bare action $\Sigma_{\rm bare}$ we obtain
\begin{eqnarray}
\Sigma_{\textrm{bare}}(\Phi_0) & = & \int d^{4}x\left\{ \frac{1}{4}F_{\mu\nu}^{a}\left(A_{0}\right)F_{\mu\nu}^{a}\left(A_{0}\right)+\lambda_{0}v_{0}^{2}h_{0}^{2}+\lambda_{0}v_{0}h_{0}^{3}+\lambda_{0}v_{0}h_{0}\rho_{0}^{a}\rho_{0}^{a}\right.\nonumber \\
 &  & +\frac{1}{4}\lambda_{0}h_{0}^{4}+\frac{1}{2}\lambda_{0}h_{0}^{2}\rho_{0}^{a}\rho_{0}^{a}+\frac{1}{4}\lambda_{0}\left(\rho_{0}^{a}\rho_{0}^{a}\right)^{2}\nonumber \\
 &  & +\frac{1}{2}\left(\partial_{\mu}h_{0}\right)^{2}+\frac{1}{2}\left(\partial_{\mu}\rho_{0}^{a}\right)^{2}\nonumber \\
 &  & +\frac{1}{2}g_{0}A_{0\mu}^{a}\rho_{0}^{a}\left(\partial_{\mu}h_{0}\right)-\frac{1}{2}g_{0}\left(v_{0}+h_{0}\right)\left(\partial_{\mu}\rho_{0}^{a}\right)A_{0\mu}^{a}+\frac{1}{2}g_{0}\varepsilon^{abc}A_{0\mu}^{a}\rho_{0}^{b}\partial_{\mu}\rho_{0}^{c}\nonumber \\
 &  & +\frac{1}{8}g_{0}^{2}A_{0\mu}^{a}A_{0\mu}^{a}\left(v_{0}+h_{0}\right)^{2}+\frac{1}{8}g_{0}^{2}A_{0\mu}^{a}A_{0\mu}^{a}\rho_{0}^{b}\rho_{0}^{b}\nonumber \\
 &  & +ib_{0}^{a}\partial_{\mu}A_{0\mu}^{a}+\overline{c}_{0}^{a}\partial_{\mu}D_{\mu}^{ab}\left(A_{0}\right)c_{0}^{b}\nonumber \\
 &  & -K_{0\mu}^{a}D_{\mu}^{ab}\left(A_{0}\right)c_{0}^{b}+L_{0}^{a}\frac{g}{2}\varepsilon^{abc}c_{0}^{b}c_{0}^{c}+H_{0}\frac{g_{0}}{2}c_{0}^{a}\rho_{0}^{a}+P_{0}^{a}\left(-\frac{g_{0}}{2}c_{0}^{a}\left(v_{0}+h_{0}\right)+\frac{g_{0}}{2}\varepsilon^{abc}c_{0}^{b}\rho_{0}^{c}\right)\nonumber \\
 &  & +J_{0}O_{0}+\eta_{0}v_{0}^{2}\nonumber \\
 &  & +\Omega_{0\mu}^{a}R_{0\mu}^{a}+\Upsilon_{0\mu}^{a}\left(-\varepsilon^{abc}\left(D_{\mu}^{bd}\left(A_{0}\right)c_{0}^{d}\right)\overline{c}_{0}^{c}+i\varepsilon^{abc}A_{0\mu}^{b}b_{0}^{c}\right)+\zeta_{0\mu}^{a}\varepsilon^{abc}A_{0\mu}^{b}\overline{c}_{0}^{c}+\Theta_{0\mu}^{a}\partial_{\mu}b_{0}^{a}\nonumber \\
 &  & +\left(Z_{\lambda}-1\right)\lambda_{0}\left(\frac{1}{4\lambda_{0}^{2}}J_{0}^{2}-\frac{1}{4\lambda_{0}}\Omega_{0\mu}^{a}\Omega_{0\mu}^{a}O_{0}-\frac{1}{8\lambda^{2}}J_{0}\Omega_{0\mu}^{a}\Omega_{0\mu}^{a}+\frac{1}{64\lambda_{0}^{2}}\Omega_{0\mu}^{a}\Omega_{0\mu}^{a}\Omega_{0\nu}^{b}\Omega_{0\nu}^{b}\right)\nonumber \\
 &  & +\delta\sigma_{0}v_{0}^{2}O_{0}\nonumber \\
 &  & +\left(\delta\sigma_{0}-\lambda_{0}\left(Z_{h}-1\right)\right)\left(-\frac{1}{2\lambda_{0}^{2}}J_{0}^{2}-\frac{1}{8\lambda_{0}}v_{0}^{2}\Omega_{0\mu}^{a}\Omega_{0\mu}^{a}+\frac{1}{4\lambda_{0}^{2}}J_{0}\Omega_{0\mu}^{a}\Omega_{0\mu}^{a}+\frac{1}{4\lambda_{0}}\Omega_{0\mu}^{a}\Omega_{0\mu}^{a}O_{0}-\frac{1}{32\lambda_{0}^{2}}\Omega_{0\mu}^{a}\Omega_{0\mu}^{a}\Omega_{0\nu}^{b}\Omega_{0\nu}^{b}\right)\nonumber \\
 &  & +\delta a_{0}\left(v_{0}^{4}+\frac{1}{\lambda_{0}^{2}}J_{0}^{2}+\frac{1}{2\lambda_{0}}\Omega_{0\mu}^{a}\Omega_{0\mu}^{a}v_{0}^{2}-\frac{1}{2\lambda_{0}^{2}}J_{0}\Omega_{0\mu}^{a}\Omega_{0\mu}^{a}+\frac{1}{16\lambda_{0}^{2}}\Omega_{0\mu}^{a}\Omega_{0\mu}^{a}\Omega_{0\nu}^{b}\Omega_{0\nu}^{b}\right)\nonumber \\
 &  & \left.-\delta\theta_{0}\mathcal{F}_{\mu\nu}^{a}\left(\Omega_{0}\right)\mathcal{F}_{\mu\nu}^{a}\left(\Omega_{0}\right)+Z_{A\Upsilon}^{\frac{1}{2}}\left(-K_{0\mu}^{a}\zeta_{0\mu}^{a}-\left(\partial_{\mu}\overline{c}_{0}^{a}\right)\zeta_{0\mu}^{a}-ib_{0}^{a}\partial_{\mu}\Upsilon_{0\mu}^{a}\right)\right\} ,\label{eq:bare_action}
\end{eqnarray}
where $\{\Phi_0\}$ is a short-hand notation for all bare fields, coupling constants, parameters and sources, while
\begin{eqnarray}
O_{0} & \coloneqq & O\left(h_{0},v_{0},\rho_{0}^{a}\right)
\end{eqnarray}
and
\begin{eqnarray}
R_{0\mu}^{a} & \coloneqq & R_{\mu}^{a}\left(A_{0\mu}^{a},h_{0},\rho_{0}^{a},v_{0},g_{0}\right).
\end{eqnarray}
Bare quantities and renormalized quantities are found to be related as follow:
\begin{eqnarray}
A_{0\mu}^{a} & = & Z_{AA}^{\frac{1}{2}}A_{\mu}^{a}+Z_{A\Upsilon}^{\frac{1}{2}}\Upsilon_{\mu}^{a}\label{eq:A_renorm}\,,\quad
h_{0} ~=~ Z_{h}^{\frac{1}{2}}h\,,\quad
\rho_{0}^{a}~=~ Z_{\rho}^{\frac{1}{2}}\rho^{a}\,,\quad
v_{0}~=~ Z_{v}^{\frac{1}{2}}v \,,\quad
c_{0}^{a}~=~Z_{c}^{\frac{1}{2}}c^{a} \,,\quad
\overline{c}_{0}^{a}~=~Z_{\overline{c}}^{\frac{1}{2}}\overline{c}^{a}\,, \nonumber\\
b_{0}^{a} & = & Z_{b}^{\frac{1}{2}}b^{a}\,,\quad
g_{0}~=~Z_{g}g\,,\quad
\lambda_{0}~=~Z_{\lambda}\lambda\,,\quad
K_{0\mu}^{a}~=~Z_{K}K_{\mu}^{a} \,,\quad
L_{0}^{a} ~=~ Z_{L}L^{a} \,,\quad
H_{0} ~=~ Z_{H}H\,, \nonumber\\ 
P_{0}^{a} &=& Z_{P}P^{a}\,,\quad \Omega_{0\mu}^{a} ~=~ Z_{\Omega}\Omega_{\mu}^{a} \,,\quad
\Upsilon_{0\mu}^{a} ~=~ Z_{\Upsilon}\Upsilon_{\mu}^{a} \,,\quad
\zeta_{0\mu}^{a} ~=~ Z_{\zeta}\zeta_{\mu}^{a} \,,\quad
\Theta_{0\mu}^{a} ~=~ Z_{\Theta}\Theta_{\mu}^{a}\,, \nonumber\\
J_{0} & = & Z_{JJ}J+Z_{J\eta}\eta \,,\quad
\eta_{0} ~=~ Z_{\eta J}J+Z_{\eta\eta}\eta \,,\quad
\delta\sigma_{0} ~=~ \epsilon\delta\sigma \,,\quad
\delta a_{0} ~=~ \epsilon\delta a \,,\quad
\delta\theta_{0} ~=~ \epsilon\delta\theta \;, \label{barequantities}
\end{eqnarray}
where $\epsilon$ is the expansion parameter. For instance, in a loop expansion in power series of $\hbar$, where the counterterm is determined recursively, order by order, the expansion parameter $\epsilon$ is nothing but the loop expansion parameter $\hbar$.

By direct inspection of the most general counterterm $\Sigma_{\rm ct}$, eqs.\eqref{delta_final_nontrivial},\eqref{eq:delta_final}, for the $Z$'s factors we have
\begin{eqnarray}
Z_{AA}^{\frac{1}{2}} & = & Z_{b}^{-\frac{1}{2}}=Z_{L}=Z_{\Theta}=1+\epsilon\frac{1}{2}\left(a_{0}-2d_{1}\right)\,,\quad
Z_{g} ~=~ 1-\epsilon\frac{1}{2}a_{0} \,,\quad
Z_{h}^{\frac{1}{2}} ~=~ Z_{\rho}^{\frac{1}{2}}=Z_{v}^{\frac{1}{2}}=Z_{\eta\eta}^{-\frac{1}{2}}=1+\epsilon d_{13}\,, \nonumber\\
Z_{A\Upsilon}^{\frac{1}{2}} & = & -\frac{1}{g}\left(Z_{AA}^{\frac{1}{2}}+Z_{g}-2\right) \,,\quad
Z_{\Omega} ~=~ Z_{\Upsilon}=1 \,,\quad
Z_{\lambda} ~=~ 1+\epsilon a_{1}\,, \nonumber\\
Z_{c}^{\frac{1}{2}} &=& Z_{\overline{c}}^{\frac{1}{2}}=Z_{K}=1+\epsilon\frac{1}{2}d_{1}=Z_{g}^{-\frac{1}{2}}Z_{AA}^{-\frac{1}{4}}\,, \nonumber\\
Z_{H} & = & Z_{P}=Z_{g}^{-\frac{1}{2}}Z_{AA}^{\frac{1}{4}}Z_{h}^{-\frac{1}{2}} \,,\quad
Z_{\zeta} ~=~ Z_{g}^{\frac{1}{2}}Z_{AA}^{-\frac{1}{4}} \,,\quad
Z_{JJ} ~=~ 1+\epsilon\left(-\frac{a_{2}}{\lambda}+a_{1}\right)\,, \nonumber\\
Z_{J\eta} & = & 0 \,,\quad
Z_{\eta J} ~=~ \epsilon\left(\frac{a_{2}}{2\lambda}-2\frac{a_{5}}{\lambda}+d_{13}\right) \,,\quad
\delta\sigma ~=~ \left(a_{2}+2\lambda d_{13}\right) \,,\quad
\delta\theta ~=~ \frac{a_{12}}{2} \,,\quad
\delta a ~=~ a_{5} \;. \label{Zfactors}
\end{eqnarray}
One notices that
$\left\{ \delta\sigma_{0},\delta a_{0},\delta\theta_{0}\right\} $
as well as  $\left\{ Z_{J\eta},Z_{\eta J}\right\} $ start from one-loop onward. In particular, the renormalization factors $\left\{ Z_{J\eta},Z_{\eta J}\right\} $ give rise to a $2 \times 2$ mixing matrix between the external sources $(J,\eta)$ coupled, respectively, to the operator $O$ and to the parameter $v^2$, a feature already observed in the case of the $U(1)$ Higgs model \cite{Capri:2020ppe,Dudal:2021pvw}.

Also, due to the introduction of the BRST exact composite operator $\left[s\left(\varepsilon^{abc}A_{\mu}^{b}\overline{c}^{c}\right)\right]$ coupled to the source $\Upsilon_{\mu}^{a}$, see eq.~\eqref{eq:exact_term}, from the expression for $A_{0\mu}^{a}$ in eqs.~\eqref{barequantities} we see that there is a mixing between the gauge field $A^a_\mu$ and the external source $\Upsilon_{\mu}^{a}$ which has precisely the same quantum numbers of $A^a_\mu$. Both $A^a_\mu$ and $\Upsilon_{\mu}^{a}$ have dimension one, vanishing ghost number and are share the property of being not BRST invariant. Let us elaborate a little bit more on this point. First, from expressions \eqref{Zfactors}, it turns out that
\begin{equation}
Z_{A\Upsilon}^{\frac{1}{2}} = - \epsilon \frac{d_1}{g}  \;, \label{mixUps}
\end{equation}
from which it follows that the mixing factor $Z_{A\Upsilon}^{\frac{1}{2}}$ starts from one-loop onward too. The existence of such a mixing means essentially that the elementary field $A^ a_\mu$ has a non-vanishing overlap with the composite operator
$\left[s\left(\varepsilon^{abc}A_{\mu}^{b}\overline{c}^{c}\right)\right]$, namely
\begin{equation}
\langle A^p_\nu(x) \left[s[ \left(\varepsilon^{abc}A_{\mu}^{b}\overline{c}^{c}\right)(y)] \right] \rangle =  \frac{\delta^2 Z_c}{\delta J^p_{A\nu}(x) \delta \Upsilon^a_\mu(y)} \Biggl|_{\rm sources=0} \neq 0 \;, \label{mixconn}
\end{equation}
where $Z_c$ is the connected generating functional, eq.~\eqref{eq:connected_funct}, and
\begin{equation}
A^p_\nu(x) = \frac{\delta Z_c}{\delta J^p_{A\nu}(x)} \;.
\end{equation}
The renormalization factor $Z_{A\Upsilon}^{\frac{1}{2}}$ would be needed to take into account the divergences present in correlation functions of the type of eq.~\eqref{mixconn}. Though, it worth to remind that the source $\Upsilon_{\mu}^{a}$ belongs to a BRST doublet:
\begin{eqnarray}
s\Upsilon_{\mu}^{a}  =  \zeta_{\mu}^{a}\nonumber \;, \quad
s\zeta_{\mu}^{a}  =  0 \;,\label{eq:doublet_sources_v1}
\end{eqnarray}
meaning that it can only enter in the exact part of the BRST cohomology. As a consequence, the composite operator $[s \left(\varepsilon^{abc}A_{\mu}^{b}\overline{c}^{c}\right)]$ has no overlap with the two local operators $(O,R^a_\mu)$ we are interested in, due to
\begin{equation}
\langle O(x_1)....O(x_n) \left( s Q(y) \right) \rangle = \langle s \left(  O(x_1)....O(x_n)  Q(y) \right) \rangle = 0\;, \label{foo}
\end{equation}
\begin{equation}
\langle R^{a_1}_{\mu_1}(x_1)....R^{a_n}_{\mu_n}(x_n) \left( s Q(y) \right) \rangle = \langle s \left(  R^{a_1}_{\mu_1}(x_1)....R^{a_n}_{\mu_n}(x_n)  Q(y) \right) \rangle = 0\;, \label{frr}
\end{equation}
for an arbitrary quantity $Q(y)$. Eqs.~\eqref{foo} and \eqref{frr}  follow from the fact that, as we have seen before in Section~\ref{3}, the operators $(O, R^a_\mu)$ are non-trivial (physical) elements of the cohomology of the BRST operator $s$ and cannot be cast in the form of an exact $s$-variation.

Taking into account that after differentiating the connected functional $Z_c$ with respect to  $(\Omega^a_\mu,J)$ all sources will be set to zero, we see that the mixing term in the external source $\Upsilon_{\mu}^{a}$ present in the bare field $A_{0\mu}^{a}$, eqs.~\eqref{barequantities}, has in fact no practical consequences on the BRST invariant correlation functions of eqs.~\eqref{2-pc}.

Let us end this section by underlining two relevant results, valid to all orders, which follow from the algebraic analysis presented here:
\begin{itemize}
\item the non-renormalization theorem \cite{Taylor:1971ff,Blasi:1990xz} of the ghost-antighost-gauge boson vertex  expressed by the relationship
\begin{eqnarray}
Z_{g}Z_{c}Z_{AA}^{\frac{1}{2}} & = & 1 \;, \label{nrlandau}
\end{eqnarray}
generalizes to the case with a fundamental Higgs field present as well. As mentioned in the introduction, this theorem is playing an important role in the study of the infrared properties of the correlation functions of non-Abelian gauge theories, see for instance \cite{Alkofer:2000wg,Binosi:2009qm,Huber:2018ned} for applications to the study of the Schwinger-Dyson equations.

\item as a consequence of the non-renormalization of the source $\Omega^a_\mu$, eqs.\eqref{Zfactors}, coupled to the gauge invariant operator $\{ R^a_\mu \}$, i.e.~
\begin{equation}
Z_{\Omega}=1\,,
\end{equation}
the anomalous dimension of $\{ R^a_\mu \}$ vanishes to all orders: $\gamma_R= \mu \partial_\mu \log(Z_\Omega)=0$ where $\mu$ stands for the renormalization scale energy. This result is in perfect agreement with the fact that $\{ R^a_\mu \}$ are the conserved Noether currents of the custodial symmetry, see Section \ref{4}.
\end{itemize}

\section{Tadpoles,  vacuum energy and the perturbative condensate $\langle O\rangle$\label{7}}
For the benefit of the reader, in this section we briefly remind a few properties related to the Ward identity  (\ref{eq:h_equation}) which, when written in terms of the $1PI$ generating functional $\Gamma$, takes the following form
\begin{eqnarray}
\int d^{4}x\left(\frac{\delta\Gamma}{\delta h}-2\lambda v\frac{\delta\Gamma}{\delta J}\right)-\frac{\partial\Gamma}{\partial v} & = & \int d^{4}xv\left(J-2\eta\right). \label{h_eq_effective_action}
\end{eqnarray}
As already emphasized in the case of the $U(1)$ Higgs model, see \cite{Capri:2020ppe,Dudal:2021pvw}, this Ward identity can be written down only when the composite operators $(O,R^a_\mu)$, eqs.\eqref{Oop},\eqref{R_op}, are introduced in the starting action from the very beginning. In particular, as shown in \cite{Dudal:2021pvw}, the Ward identity \eqref{h_eq_effective_action} has quite nice consequences on the vacuum energy ${\cal E}_v$ of the theory. In fact, setting all sources to zero, one gets
\begin{equation}
\frac{\partial {\cal E}_v}{\partial v} = \langle h \rangle - 2 \lambda v \langle O \rangle \;, \label{ven}
\end{equation}
implying a relationship between the vacuum energy ${\cal E}_v$, the tadpoles $\langle h \rangle$ and the dimension two condensate $\langle O \rangle$. Notice that the condensate here is not necessarily the perturbative one, i.e.~the relation \eqref{ven} is exact.

At the perturbative level, as we have seen, the most general local non-trivial invariant counterterm, eq.~\eqref{delta_final_nontrivial}, contains the two  BRST invariant counterterms  $(a_{2}v^{2}O)$ and $(a_5 v^4)$, where $a_2$ and $a_5$ are free parameters which start from one-loop onward. The presence of the counterterm $a_{2}v^{2}O$ is a well known property of both $U(1)$ and $SU(2)$ Higgs models \cite{Becchi:1975nq,Kraus:1998ud}. The parameter  $a_2$ can be chosen so as to ensure the vanishing of the tadpoles, $ \langle h \rangle =0$, order by order in the loop expansion. On the other hand, as discussed in \cite{Dudal:2021pvw}, the free coefficient $a_5$ can be  fixed by requiring the vanishing to all orders of the perturbative dimension two condensate, namely $\langle O \rangle_{\rm pert}=0$. Therefore, the Ward identity \eqref{ven} ensures that the perturbative vacuum energy keeps its minimum at $v$ during the renormalization process:
\begin{equation}
\frac{\partial {\cal E}_{v, {\rm pert}}}{\partial v} = 0  \;. \label{minen}
\end{equation}
We highlight that eq.~\eqref{minen} follows from a Ward identity, eq.~\eqref{h_eq_effective_action}, which can be written down only when the composite operators $(O, R^a_\mu)$ are taken into account in the starting action. To some extent, the gauge invariant setup for the Higgs particle and the gauge vector boson provided by $(O,R^a_\mu)$ gives us a very nice way to check out, by means of the Ward identity \eqref{ven}, that the perturbative vacuum energy ${\cal E}_{v, {\rm pert}}$ displays the desired property of attaining its minimum at $v$.  Once the perturbative setup is settled in this fashion, it leads to the interesting question how the potential generation of a non-perturbative condensate $\langle O \rangle_{\rm non-pert}$ would influence the dynamics, including the new vacuum, since enforced by the exact identity \eqref{ven}, this will shift the minimum configuration. We hope to come back to this issue in future work.

\section{The longitudinal component of the two-point correlation function  of $\left\langle R_{\mu}^{a}\left(x\right)R_{\nu}^{b}\left(y\right)\right\rangle $ \label{8}}
This section is devoted to the study of the consequences stemming from the custodial Ward identity, eq.~\eqref{eq:custodial_eq}, on the longitudinal component of the two-point correlation function  of $\left\langle R_{\mu}^{a}\left(x\right)R_{\nu}^{b}\left(y\right)\right\rangle $.
At the quantum level, the Ward identity \eqref{eq:custodial_eq} reads
\begin{eqnarray}
\mathcal{C}^{n}\left(\Gamma\right) & = & \frac{1}{4}gv^{2}\partial_{\mu}\Omega_{\mu}^{n}+i\varepsilon^{mnp}\partial_{\mu}\left(\Upsilon_{\mu}^{m}b^{p}\right)+\varepsilon^{mnp}\partial_{\mu}\left(\zeta_{\mu}^{m}\overline{c}^{p}\right)-g\varepsilon^{mnp}\partial_{\mu}\left(b^{p}\Theta_{\mu}^{m}\right)-i\partial^{2}b^{n} \;. \label{eq:custodial_gamma}
\end{eqnarray}
Moving to the connected generating functional $Z_c$, eq.~\eqref{eq:connected_funct}, one obtains
\begin{eqnarray}
&  & -g\varepsilon^{mnp}\left(J_{A}\right)_{\alpha}^{p}\frac{\delta Z_{c}}{\delta\left(J_{A}\right)_{\alpha}^{m}}-g\varepsilon^{mnp}\left(J_{\rho}\right)^{p}\frac{\delta Z_{c}}{\delta\left(J_{\rho}\right)^{m}}-g\varepsilon^{mnp}\left(J_{c}\right)^{p}\frac{\delta Z_{c}}{\delta\left(J_{c}\right)^{m}}\nonumber \\
 &  & -g\varepsilon^{mnp}\left(J_{\overline{c}}\right)^{p}\frac{\delta Z_{c}}{\delta\left(J_{\overline{c}}\right)^{m}}-g\varepsilon^{mnp}\left(J_{b}\right)^{p}\frac{\delta Z_{c}}{\delta\left(J_{b}\right)^{m}} +g\varepsilon^{mnp}K_{\mu}^{p}\frac{\delta Z_{c}}{\delta K_{\mu}^{m}}+g\varepsilon^{mnp}L^{p}\frac{\delta Z_{c}}{\delta L^{m}}+g\varepsilon^{mnp}P^{p}\frac{\delta Z_{c}}{\delta P^{m}}\nonumber \\
 &  & +g\varepsilon^{mnp}\Omega_{\mu}^{p}\frac{\delta Z_{c}}{\delta\Omega_{\mu}^{m}}+g\varepsilon^{mnp}\Upsilon_{\mu}^{p}\frac{\delta Z_{c}}{\delta\Upsilon_{\mu}^{m}}+g\varepsilon^{mnp}\zeta_{\mu}^{p}\frac{\delta Z_{c}}{\delta\zeta_{\mu}^{m}}+g\varepsilon^{mnp}\Theta_{\mu}^{p}\frac{\delta Z_{c}}{\delta\Theta_{\mu}^{m}}\nonumber \\
 &  & -\partial_{\mu}\left(J_{A}\right)_{\mu}^{n}-g\partial_{\mu}\frac{\delta Z_{c}}{\delta\Omega_{\mu}^{n}}-\frac{1}{2}g\partial_{\mu}\left(\Omega_{\mu}^{n}\frac{\delta Z_{c}}{\delta J}\right)+g\partial_{\mu}\frac{\delta Z_{c}}{\delta\Upsilon_{\mu}^{n}}\nonumber \\
 & = & \frac{1}{4}gv^{2}\partial_{\mu}\Omega_{\mu}^{n}+i\varepsilon^{mnp}\partial_{\mu}\left(\Upsilon_{\mu}^{m}\frac{\delta Z_{c}}{\delta\left(J_{b}\right)^{p}}\right)+\varepsilon^{mnp}\partial_{\mu}\left(\zeta_{\mu}^{m}\frac{\delta Z_{c}}{\delta\left(J_{\overline{c}}\right)^{p}}\right)-g\varepsilon^{mnp}\partial_{\mu}\left(\Theta_{\mu}^{m}\frac{\delta Z_{c}}{\delta\left(J_{b}\right)^{p}}\right)-i\partial^{2}\frac{\delta Z_{c}}{\delta\left(J_{b}\right)^{n}}\label{eq:custodial_connected}
\end{eqnarray}
where $\left(J_{A}\right)_{\mu}^{a}$, $\left(J_{\rho}\right)^{a}$,
$\left(J_{c}\right)^{a}$, $\left(J_{\overline{c}}\right)^{a}$ and
$\left(J_{b}\right)^{a}$ are the sources of $A_{\mu}^{a}$, $\rho^{a}$,
$c^{a}$ and $\overline{c}^{a}$, respectively. Acting with $\delta/\delta\Omega_{\nu}^{\ell}$
on (\ref{eq:custodial_connected})  and setting all
sources to zero, we find
\begin{eqnarray*}
g\partial_{\mu}^{x}\left\langle R_{\nu}^{\ell}\left(y\right)R_{\mu}^{n}\left(x\right)\right\rangle -\frac{1}{2}g\delta^{n\ell}\partial_{\nu}^{x}\left(\delta\left(x-y\right)\left\langle O\right\rangle_{\rm pert}\right) & = & \frac{1}{4}gv^{2}\delta^{n\ell}\partial_{\nu}^{x}\delta\left(x-y\right)+i\left(\partial^{x}\right)^{2}\left\langle R_{\nu}^{\ell}\left(y\right)b^{n}\left(x\right)\right\rangle \;.
\end{eqnarray*}
This result can be simplified even more, since:
\begin{itemize}
\item $\left\langle R_{\nu}^{\ell}\left(y\right)b^{n}\left(x\right)\right\rangle =-i\left\langle s\left[R_{\nu}^{\ell}\left(y\right)\overline{c}^{n}\left(x\right)\right]\right\rangle =0$, due to the exact BRST invariance of the theory,
\item as we have seen in the previous section, we can adjust the vacuum counterterm $a_5 v^4$ so as to ensure that the dimension two condensate $\langle O\rangle_{\rm pert}$ vanishes order by order: $\langle O\rangle_{\rm pert}= 0 $ \;.
\end{itemize}
As a consequence, we get the important result that
\begin{eqnarray}
\partial_{\mu}^{x}\left\langle R_{\nu}^{\ell}\left(y\right)R_{\mu}^{n}\left(x\right)\right\rangle _{c} & = & \frac{1}{4}v^{2}\delta^{n\ell}\partial_{\mu}^{x}\delta\left(x-y\right)
\end{eqnarray}
\begin{eqnarray}
\Rightarrow\mathcal{L}_{\mu\nu}\left\langle R_{\nu}^{l}\left(y\right)R_{\mu}^{n}\left(x\right)\right\rangle _{c} & = & \frac{1}{4}v^{2}\delta^{nl}\;, \label{longtwo}
\end{eqnarray}
where $\mathcal{L}_{\mu\nu}=\frac{\partial_{\mu}\partial_{\nu}}{\partial^{2}}$
stands for the longitudinal projector.  Eq.~\eqref{longtwo} states that the longitudinal component of the two-point correlation function  of $\left\langle R_{\mu}^{a}\left(x\right)R_{\nu}^{b}\left(y\right)\right\rangle $ does not receive any quantum correction beyond its tree level contribution which, moreover, is fully momentum independent. This means that the longitudinal component of the gauge invariant composite operator $R^a_\mu$ cannot be associated to any propagating physical mode. Only the transverse component of $\{ R^a_\mu \}$ matters. Eq.~\eqref{longtwo} yields a non-trivial consistency check of the usefulness of the conserved  operator $R^a_\mu$ to provide a gauge invariant picture for the massive vector bosons.

\section{Conclusion \label{conclusion}}
In this work, we have studied the renormalization properties of two BRST invariant local operators, the scalar $O(x)$ and the vector $R_{\mu}(x)$, which were first introduced in a previous work \cite{Dudal:2019pyg}. These operators provide a BRST-invariant framework to describe the Higgs particle and the gauge vector boson in the $SU(2)$ Higgs model. The renormalization properties of $(O(x), R_{\mu}(x))$ were studied by introducing their respective external sources $(J(x),\Omega (x))$ in the starting action and analyzing the renormalization properties by means of
the Algebraic Renormalization framework. As could be seen in Section \ref{6}, the composite BRST-invariant description of the physical degrees of freedom of the $SU(2)$ Higgs model is renormalizable to all orders in perturbation theory.

By choosing the Landau gauge, the system enjoys the existence of a large set of Ward identities, which ends up to strongly restrict the set of free parameters needed to renormalize the theory. One especially interesting Ward identity is connected to the custodial symmetry, as discussed in Section \ref{4}. This symmetry is a generalization of the custodial symmetry discussed in the
$U(1)$ Higgs case, \cite{Dudal:2019pyg}. The particular construction of the vector operators $R^a_{\mu}$ in Section \ref{3} means that these operators are the conserved currents of the custodial symmetry. As a consequence, the corresponding source terms $\Omega^a_\mu$ does not receive any quantum corrections, i.e.~$Z_{\Omega}=1$.  This is a very powerful result, as it is rooted in a Ward identity, which should also hold non-perturbatively. Our findings here could also be of relevance to the lattice study of the gauge invariant vector operators and the corresponding part of the spectrum they describe. In general, to extract physical information, one should properly renormalize lattice correlation functions of gauge invariant operators, cf.~\cite{Costa:2021iyv} for a recent account and references. Although we have now established the non-renormalizabilit in the continuum, due to the discretization on a lattice, (finite) renormalizations might still be necessary \cite{Hatton:2019gha}, but the conserved nature of the current should also be of help here.

Another consequence of the custodial symmetry is found in Section \ref{8}, where we have investigated the longitudinal part of the correlation function $\langle R^a_{\mu}(x)R^b_{\nu}(y)\rangle$. In line with the conclusions on the one-loop perturbative corrections found in \cite{Dudal:2020uwb}, we conclude that the longitudinal part is tree level exact.

In Section \ref{7} the consequences of introducing the local gauge invariant composite operator $O(x)$ from the beginning were discussed. Besides the custodial symmetry, the $h$ equation \eqref{eq:h_equation} plays a central role in the renormalization of $O$. Particularly, we found that the number of free parameters needed to renormalize the scalar sector is the same as in the Abelian-Higgs model, (i.e.~as in the absence of $O$). Furthermore, at each perturbative order, the symmetry \eqref{h_eq_effective_action} establishes a connection between the vacuum energy $\cal E$, the $\langle O(x) \rangle$ and $\langle h(x) \rangle$.  In particular, at any order of perturbation theory, one may choose a suitable vacuum configuration and renormalization scheme such that $\langle O \rangle_{\rm pert} = 0$ and $\langle h(x) \rangle = 0$. It will be interesting to investigate the interplay of the exact Ward identity \eqref{h_eq_effective_action} and a potential non-perturbative condensate $\langle O \rangle_{\rm non-pert}$, which will influence both the vacuum energy and correlation functions.

Finally, in \cite{Dudal:2021pvw}, in the $U(1)$ case, we also set first steps in explicitly rewriting the effective action in terms of the newly defined gauge invariant operators, via means of a non-trivial path integral transformation. This amounts to considering the Equivalence Theorem, \cite{Bergere:1975tr,Haag:1958vt,Kamefuchi:1961sb,Lam:1973qa,Blasi:1998ph}, which we reinterpreted in terms of an extended (constraint) BRST cohomology. Albeit that the full scope of this needs to be established still even in the $U(1)$ case, one can already speculate that something similar, albeit more complicated, should also work out for the non-Abelian case, paving the way towards a potentially novel, explicitly gauge invariant and renormalizable, scheme to deal with quantum gauge field theories with a Higgs mechanism.

\section*{Acknowledgments}
The Conselho Nacional de Desenvolvimento Cient\'{\i}fico e Tecnol\'{o}gico (CNPq-Brazil), the Funda\c{c}\~{a}o de Amparo \`{a} Pesquisa do Estado do Rio de Janeiro (FAPERJ) and the SR2-UERJ are gratefully acknowledged for financial support. S. P. Sorella is a level PQ-1 researcher under the program Produtividade em Pesquisa-CNPq, 301030/2019-7. I. F. Justo acknowledges CAPES for the financial support under the project grant $88887.357904/2019-00$. D.M. van Egmond was partly financed by KU Leuven with a visiting researcher fellowship.


\begin{thebibliography}{100}

\bibitem{Dudal:2020uwb}
D.~Dudal, D.~M.~van Egmond, M.~S.~Guimaraes, L.~F.~Palhares, G.~Peruzzo and S.~P.~Sorella,
Eur. Phys. J. C \textbf{81}, no.3, 222 (2021)
[arXiv:2008.07813 [hep-th]].


\bibitem{Dudal:2019pyg}
D.~Dudal, D.~M.~van Egmond, M.~S.~Guimaraes, O.~Holanda, L.~F.~Palhares, G.~Peruzzo and S.~P.~Sorella,
JHEP {\bf 2002}, 188 (2020)
[arXiv:1912.11390 [hep-th]].

\bibitem{tHooft:1980xss}
G.~'t Hooft, C.~Itzykson, A.~Jaffe, H.~Lehmann, P.~K.~Mitter, I.~M.~Singer and R.~Stora,
NATO Sci. Ser. B \textbf{59}, pp.1-438 (1980).


\bibitem{fms1}
J.~Frohlich, G.~Morchio and F.~Strocchi,
Phys. Lett. B \textbf{97}, 249-252 (1980).


\bibitem{fms2}
J.~Frohlich, G.~Morchio and F.~Strocchi,
Nucl. Phys. B \textbf{190}, 553-582 (1981).


\bibitem{Binosi:2009qm}
D.~Binosi and J.~Papavassiliou,
Phys. Rept. \textbf{479}, 1-152 (2009)
[arXiv:0909.2536 [hep-ph]].


\bibitem{Maas:2017wzi}
A.~Maas,
Prog. Part. Nucl. Phys. \textbf{106}, 132-209 (2019)
[arXiv:1712.04721 [hep-ph]].

\bibitem{Maas:2017xzh}
A.~Maas, R.~Sondenheimer and P.~T\"orek,
Annals Phys. \textbf{402}, 18-44 (2019)
[arXiv:1709.07477 [hep-ph]].

\bibitem{Maas:2018xxu}
A.~Maas and P.~T\"orek,
Annals Phys. \textbf{397}, 303-335 (2018)
doi:10.1016/j.aop.2018.08.018
[arXiv:1804.04453 [hep-lat]].

\bibitem{Maas:2020kda}
A.~Maas and R.~Sondenheimer,
Phys. Rev. D \textbf{102}, 113001 (2020)
[arXiv:2009.06671 [hep-ph]].

\bibitem{Sondenheimer:2019idq}
R.~Sondenheimer,
Phys. Rev. D \textbf{101}, no.5, 056006 (2020)
[arXiv:1912.08680 [hep-th]].


\bibitem{Struyve:2011nz}
W.~Struyve,
Stud. Hist. Phil. Sci. B \textbf{42}, 226-236 (2011)
[arXiv:1102.0468 [physics.hist-ph]].

\bibitem{Berghofer:2021ufy}
P.~Berghofer, J.~Fran\c{c}ois, S.~Friederich, H.~Gomes, G.~Hetzroni, A.~Maas and R.~Sondenheimer,
[arXiv:2110.00616 [physics.hist-ph]].


\bibitem{Higgs:1966ev}
P.~W.~Higgs,
Phys. Rev. \textbf{145}, 1156-1163 (1966).



\bibitem{Nielsen:1975fs}
N.~K.~Nielsen,
Nucl. Phys. B \textbf{101}, 173-188 (1975).

\bibitem{Gambino:1998ec}
P.~Gambino, P.~A.~Grassi and F.~Madricardo,
Phys. Lett. B \textbf{454}, 98-104 (1999)
[arXiv:hep-ph/9811470 [hep-ph]].

\bibitem{Gambino:1999ai}
P.~Gambino and P.~A.~Grassi,
Phys. Rev. D \textbf{62}, 076002 (2000)
[arXiv:hep-ph/9907254 [hep-ph]].

\bibitem{weinberg}
S.~Weinberg, ``The Quantum Theory of Fields. Vol.~1: Foundations'', Cambridge University Press (2013).



\bibitem{Dudal:2019aew}
D.~Dudal, D.~M.~van Egmond, M.~S.~Guimar\~aes, O.~Holanda, B.~W.~Mintz, L.~F.~Palhares, G.~Peruzzo and S.~P.~Sorella,
Phys. Rev. D \textbf{100}  no.6, 065009 (2019)
[arXiv:1905.10422 [hep-th]].

\bibitem{Dudal:2019pyg}
D.~Dudal, D.~M.~van Egmond, M.~S.~Guimaraes, O.~Holanda, L.~F.~Palhares, G.~Peruzzo and S.~P.~Sorella,
JHEP \textbf{02}, 188 (2020)
[arXiv:1912.11390 [hep-th]].

\bibitem{Capri:2020ppe}
M.~A.~L.~Capri, I.~F.~Justo, L.~F.~Palhares, G.~Peruzzo and S.~P.~Sorella,
Phys. Rev. D \textbf{102}  no.3, 033003 (2020)
[arXiv:2007.01770 [hep-th]].

\bibitem{Dudal:2021pvw}
D.~Dudal, G.~Peruzzo and S.~P.~Sorella,
JHEP \textbf{10}, 039 (2021)
[arXiv:2105.11011 [hep-th]].

\bibitem{Itzykson:1980rh}
 C.~Itzykson and J.~B.~Zuber,
``Quantum Field Theory,'' New York, USA: McGraw-Hill (1980).



\bibitem{Becchi:1975nq}
C.~Becchi, A.~Rouet and R.~Stora,
Annals Phys. \textbf{98}, 287-321 (1976).

\bibitem{Becchi:1974md}
C.~Becchi, A.~Rouet and R.~Stora,
Commun. Math. Phys. \textbf{42}, 127-162 (1975).

\bibitem{Becchi:1974xu}
C.~Becchi, A.~Rouet and R.~Stora,
Phys. Lett. B \textbf{52}, 344-346 (1974).



\bibitem{Piguet:1995er}
O.~Piguet and S.~P.~Sorella,
``Algebraic renormalization: Perturbative renormalization, symmetries and anomalies,''
Lect. Notes Phys. Monogr. \textbf{28}, 1-134 (1995).


\bibitem{Taylor:1971ff}
J.~C.~Taylor,
Nucl. Phys. B \textbf{33}, 436-444 (1971).


\bibitem{Blasi:1990xz}
A.~Blasi, O.~Piguet and S.~P.~Sorella,
Nucl. Phys. B \textbf{356}, 154-162 (1991).


\bibitem{Alkofer:2000wg}
R.~Alkofer and L.~von Smekal,
Phys. Rept. \textbf{353}, 281 (2001).
[arXiv:hep-ph/0007355 [hep-ph]].

\bibitem{Binosi:2009qm}
D.~Binosi and J.~Papavassiliou,
Phys. Rept. \textbf{479}, 1-152 (2009)
[arXiv:0909.2536 [hep-ph]].

\bibitem{Huber:2018ned}
M.~Q.~Huber,
Phys. Rept. \textbf{879}, 1-92 (2020)
[arXiv:1808.05227 [hep-ph]].

\bibitem{Kraus:1998ud}
E.~Kraus,
Acta Phys. Polon. B \textbf{29}, 2647-2654 (1998)
[arXiv:hep-th/9807102 [hep-th]].


\bibitem{Costa:2021iyv}
M.~Costa, I.~Karpasitis, T.~Pafitis, G.~Panagopoulos, H.~Panagopoulos, A.~Skouroupathis and G.~Spanoudes,
Phys. Rev. D \textbf{103} no.9, 094509 (2021)
[arXiv:2102.00858 [hep-lat]].

\bibitem{Hatton:2019gha}
D.~Hatton \textit{et al.} [HPQCD],
Phys. Rev. D \textbf{100}, no.11, 114513 (2019)
[arXiv:1909.00756 [hep-lat]].


\bibitem{Bergere:1975tr}
M.~C.~Bergere and Y.~M.~P.~Lam,
Phys. Rev. D \textbf{13}, 3247-3255 (1976).

\bibitem{Haag:1958vt}
R.~Haag,
Phys. Rev. \textbf{112}, 669-673 (1958).

\bibitem{Kamefuchi:1961sb}
S.~Kamefuchi, L.~O'Raifeartaigh and A.~Salam,
Nucl. Phys. \textbf{28}, 529-549 (1961).

\bibitem{Lam:1973qa}
Y.~M.~P.~Lam,
Phys. Rev. D \textbf{7}, 2943-2949 (1973).

\bibitem{Blasi:1998ph}
A.~Blasi, N.~Maggiore, S.~P.~Sorella and L.~C.~Q.~Vilar,
Phys. Rev. D \textbf{59}, 121701 (1999)
[arXiv:hep-th/9812040 [hep-th]].


\end{thebibliography}
\end{document}